\documentclass[12pt]{article}

\usepackage{fullpage}
\usepackage{amsmath,amsfonts,amsthm,amssymb}
\usepackage{makeidx}
\usepackage{subfigure}
\usepackage{setspace}
\usepackage{Tabbing}
\usepackage{fancyhdr}
\usepackage{lastpage}
\usepackage{extramarks}
\usepackage{chngpage}
\usepackage{soul,color}
\usepackage{graphicx,float,wrapfig}
\usepackage{sidecap}
\usepackage[colorlinks,hyperindex]{hyperref}
\hypersetup{citecolor=Magenta, linkcolor=Blue, urlcolor=Red}
 \usepackage[usenames,dvipsnames]{pstricks}
\usepackage{epsfig}
 \usepackage{pst-grad} 
 \usepackage{pst-plot} 
\input{epsf}

\def\rf#1{(\ref{eq:#1})}
\def\lab#1{\label{eq:#1}}

\def\br{\begin{eqnarray}}
\def\er{\end{eqnarray}}
\def\be{\begin{equation}}
\def\ee{\end{equation}}
\def\({\left(}
\def\){\right)}
\def\rlx{\relax\leavevmode}
\def\IR{\rlx\hbox{\rm I\kern-.18em R}}

	\newcommand{\pa}{\partial}

	\newcommand{\ba}{\begin{array}}
	\newcommand{\ea}{\end{array}}
	\newcommand{\beqa}{\begin{equation}\begin{array}{rcl}}
	\newcommand{\eeqa}[1]{\end{array}\label{#1}\end{equation}}

	\renewcommand{\=}{&=&} 

\def\IZ{\rlx\hbox{\sf Z\kern-.4em Z}}
\def\IR{\rlx\hbox{\rm I\kern-.18em R}}
\def\IC{\rlx\hbox{\,$\inbar\kern-.3em{\rm C}$}}
\def\one{\hbox{{1}\kern-.25em\hbox{l}}}

\newcommand{\lp}{\left(}\newcommand{\rp}{\right)}
\newcommand{\lc}{\left[}\newcommand{\rc}{\right]}

\newcommand{\sbr}[2]{\left\lbrack\,{#1}\, ,\,{#2}\,\right\rbrack}

\def\IZ{\rlx\hbox{\sf Z\kern-.4em Z}}
\def\IR{\rlx\hbox{\rm I\kern-.18em R}}
\def\IC{\rlx\hbox{\,$\inbar\kern-.3em{\rm C}$}}
\def\one{\hbox{{1}\kern-.25em\hbox{l}}}

\begin{document}

\begin{titlepage}
\vspace*{-1cm}

\vskip 3cm

\vspace{.2in}
\begin{center}
{\large\bf Gauge and Integrable Theories in Loop Spaces}
\end{center}

\vspace{.5cm}

\begin{center}
L. A. Ferreira\footnote{e-mail: laf@ifsc.usp.br} and G. Luchini\footnote{e-mail: gabriel.luchini@gmail.com}

\vspace{.3 in}
\small

\par \vskip .2in \noindent
Instituto de F\'\i sica de S\~ao Carlos; IFSC/USP;\\
Universidade de S\~ao Paulo  \\ 
Caixa Postal 369, CEP 13560-970, S\~ao Carlos-SP, Brazil\\

\normalsize
\end{center}


\begin{abstract}
We propose an integral formulation of the equations of motion of a large class of field theories which leads in a quite natural and direct way to the construction of conservation laws. The approach is based on generalized non-abelian Stokes theorems for $p$-form connections, and its appropriate mathematical language is that of loop spaces. The equations of motion are written as the equality of an hyper-volume ordered  integral to an hyper-surface ordered  integral on the border of that hyper-volume. The approach applies to integrable field theories in $(1+1)$ dimensions, Chern-Simons theories in $(2+1)$ dimensions, and non-abelian gauge theories in $(2+1)$ and $(3+1)$ dimensions.  The results presented in this paper are relevant for the understanding of global properties of those theories. 
 \end{abstract} 
\end{titlepage}

\section{Introduction}
\label{sec:intro}
\setcounter{equation}{0}

Symmetries play a central role in the understanding of physical phenomena. The laws governing the fundamental interactions in gauge theories and general relativity are strongly based on symmetry principles. On the other hand the developments of non-perturbative methods to study strongly coupled system rely on symmetries revealed by deep structures like the weak-strong coupling dualities in gauge theories.  Even though the Noether symmetries of Lagrangians and equations of motion are very important in many aspects of a given theory, it is perhaps correct to say that the hidden symmetries are the ones that have proved to be most efficient in the development of exact methods for non-linear and non-perturbative phenomena. The best examples of that are low dimensional theories with applications in many areas of physics like condensed matter, integrable field theories and solitons.  The hidden symmetries responsible for the solvability of those $(1+1)$-dimensional theories appear in general as gauge symmetries of an auxiliary flat one-form connection $A_{\mu}$. In fact, the connection is a functional of the physical fields, and the zero curvature condition for  $A_{\mu}$ is equivalent to the classical equations of motion of the theory.  The crucial fact here is that the flatness condition imply that the path ordered integral of the connection between two given points is independent of the choice of the  path joining them.  That statement is a conservation law, and the conserved quantities are given by  the eigenvalues of the operator obtained by the path ordered integral of $A_{\mu}$ over the entire one dimensional space sub-manifold. The exact developments in soliton theories and in many non-linear phenomena in low dimensions, over the last decades,  were a direct consequence of that very important and simple fact. 

If higher dimensional theories present similar structures or not is an open and interesting problem. There have been several approaches to tackle that question, and we want to discuss here one which is a quite straightforward generalization of the ideas described above. One should expect the conserved quantities of a $d+1$ dimensional theory to be associated to integrals of quantities on the $d$-dimensional space sub-manifold.   However, that space can be seen as a path in a generalized loop space in the following way. Choose a reference point $x_R$ in the $(d+1)$-dimensional space-time $M$, and defined the space of maps $LM$ from the $(d-1)$-dimensional sphere $S^{d-1}$ into  $M$, such that the north pole of $S^{d-1}$ is always mapped into $x_R$, i.e. $LM=\{ \gamma:  S^{d-1} \rightarrow M \mid \gamma(0)=x_R\}$. The images of those maps are $(d-1)$-dimensional closed hyper-surfaces $\Sigma$ in $M$ based at $x_R$, and each one of them corresponds to a point of $LM$. Given a $d$-dimensional hyper-volume in $M$, one can scan it with a collection of those closed hyper-surfaces $\Sigma$. Such collection is a path in $LM$, and so the hyper-volume in $M$ can be seen  as a path in the loop space $LM$. The idea now is, for a given theory in $M$, to look for a  one-form connection ${\cal A}$ in $LM$, such that the conditions for its curvature to vanish are equivalent to the classical equations of motion of that physical theory. The flatness condition for ${\cal A}$ implies that its path ordered integral between two given points in $LM$ (hyper-surfaces in $M$) is independent of the choice of path (hyper-volume in $M$) joining them. That would lead, in a similar way to $(1+1)$-dimensional  theories, to conserved quantities as the eigenvalues of the path ordered integral of ${\cal A}$ over the paths corresponding to the $d$-dimensional space sub-manifold. That is the approach put forward in \cite{afs} and implemented in several examples of field theories in $(d+1)$ dimensions.  See \cite{afs-review} for a review of the interesting results obtained. Among  the difficulties of the approach are those associated to the non-locality  and to the reparameterization  invariance of the physical quantities. Note that a given hyper-volume in $M$ corresponds in fact to an infinite number of paths in $LM$, which is a consequence of the infinity of ways of scanning it with hyper-surfaces. So, the physical phenomena should not depend upon the change of scanning. Despite those difficulties it was possible to impose local conditions in $M$ which lead to the vanishing of the curvature of the connection ${\cal A}$ in the loop space $LM$, and made the physical quantities reparameterization invariant \cite{afs,afs-review}. 

In this paper we want to use the very same ideas proposed in \cite{afs,afs-review} to construct conserved quantities for a large class of theories, which include integrable field theories in $(1+1)$ dimensions, Chern-Simons theories with sources in $(2+1)$ dimensions, and Yang-Mills theories in $(2+1)$ and $(3+1)$ dimensions. However, instead of looking for a connection in loop space which zero curvature condition is equivalent to the classical equations of motion, we propose an integral form of those equations, related to generalizations of the non-abelian Stokes theorem, and which lead in a quite simple and direct way to the conservation laws. Consider a physical theory in a $(d+1)$-dimensional simply connected space-time $M$, and let $\Omega$ be any (in some sense topologically trivial) $d$-dimensional hyper-volume in $M$, and suppose that the dynamics of such theory can be described by integral equations of the form  
\begin{equation}
\lab{stokes}
P_{d-1} e^{\int_{\partial \Omega} {\cal F}}=P_d\,e^{\int_{\Omega} {\cal J}}
\end{equation}
where  $\partial \Omega$ is the border of $\Omega$, and where $P_{d-1}$ and $P_{d}$ stand for hyper-surface and hyper-volume ordering integrations respectively. The quantities ${\cal F}$ and ${\cal J}$ are built out of $d-1$ and $d$ forms in $M$ respectively, and which are functionals of the physical fields.  The details of the construction will be given  in the examples discussed in the next sections. However, we deal with local field theories and the equations \rf{stokes} are a direct consequence of the local differential equations  of motion of the theory and of some generalization of the non-abelian Stokes theorem. On the other hand, since \rf{stokes} is valid on any hyper-volume $\Omega$, it turns out that \rf{stokes} imply those local differential equations when $\Omega$ is taken to be infinitesimally small. In order to define the ordered integrations in \rf{stokes} we scan $\Omega$ with $(d-1)$-dimensional closed hyper-surfaces based on a reference point $x_R$ on its border  $\partial \Omega$. Therefore, the equations \rf{stokes} are not really defined on each $\Omega$, but on the generalized loop space $L\Omega=\{ \gamma:  S^{d-1} \rightarrow \Omega \mid \gamma(0)=x_R\}$, i.e. the space of mappings from the $(d-1)$-dimensional sphere $S^{d-1}$ to $\Omega$, such that its north pole is always mapped into $x_R$. Consequently, $\Omega$ can be seen as a path in  $L\Omega$, and the r.h.s. of \rf{stokes} is defined on such a path, with the l.h.s. of \rf{stokes} being evaluated on its end points. But  there is an infinite number of paths in $L\Omega$ corresponding to the same $\Omega$. When one changes the choice of path representing $\Omega$, both sides of \rf{stokes} change.  However, the non-abelian Stokes theorem leading to \rf{stokes} guarantees that the changes are such that both sides of \rf{stokes} remain equal. Therefore, \rf{stokes} transforms ``covariantly'' under change of parameterization of $\Omega$.  In addition, we show in the next sections that \rf{stokes} transforms covariantly under gauge transformations associated to the differential forms leading to the quantities  ${\cal F}$ and ${\cal J}$.

An important consequence of the integral form of the equations of motion \rf{stokes} is that if one considers a closed hyper-volume $\Omega_c$, i.e. without border, then the l.h.s of \rf{stokes} becomes trivial and one gets that 
\be
P_d\,e^{\int_{\Omega_c} {\cal J}} =\one
\lab{nicerel}
\ee
Note that $\Omega_c$ corresponds to a closed path in $L\Omega_c$. Then let us choose an intermediate point  on that path, i.e. a closed $(d-1)$-dimensional hyper-surface $\Sigma$, such that  $\Omega_c=\Omega_1+\Omega_2$, with $\Omega_1$ being the part of $\Omega_c$ going from the infinitesimal hyper-surface $\Sigma_R$ around the reference point $x_R$ to $\Sigma$, and $\Omega_2$ to the part going from $\Sigma$ back to $\Sigma_R$. Then, the ordered integration implies that $P_d\,e^{\int_{\Omega_2} {\cal J}}\, P_d\,e^{\int_{\Omega_1} {\cal J}} =\one$. The order may be reversed depending upon the definition of the ordered integration. By reverting the sense of integration along the path one gets the inverse operator, and so one can rewrite that relation as $P_d\,e^{\int_{\Omega_1} {\cal J}}= P_d\,e^{\int_{\Omega_2^{-1}} {\cal J}} $, with $\Omega_2^{-1}$ being the path $\Omega_2$ in reversed order. Since that is valid for any closed path passing through $\Sigma_R$ and $\Sigma$, one concludes that the operator $P_d\,e^{\int_{\Omega} {\cal J}}$ is independent of the path $\Omega$ joining  $\Sigma_R$ and $\Sigma$. That path independency is a conservation law, and by choosing appropriate boundary conditions as we explain in the next sections, one gets that the conserved charges are the eigenvalues of the operator obtained by the path ordered integral $P_d\,e^{\int_{\Omega} {\cal J}}$, with $\Omega$ corresponding to the whole space sub-manifold. In the examples we discuss such conserved charges are shown to be gauge invariant, and independent of the parameterization of the hyper-volumes as well as of the choice of the reference point $x_R$. In the case where the space-time is of the form ${\cal S}\times \IR$, with $\IR$ being the time, and ${\cal S}$ being a space sub-manifold without  border, i.e $\partial {\cal S}=0$, then one gets from \rf{nicerel} that $P_d\,e^{\int_{{\cal S}} {\cal J}}=\one$. Therefore, such operator is not only constant in time but trivial. In many cases, that topological property of the space-time leads to quantization of charges. That is a very important consequence of our construction.  

Even though we have not introduced a one-form connection in the loop space $L\Omega$, that concept is hidden in the quantity ${\cal J}$. In addition, since we use generalizations of the non-abelian Stokes theorem, the quantity  ${\cal J}$ corresponds to some sort of curvature of the quantity ${\cal F}$, now seen as a connection on a lower loop space $L\partial\Omega$, made of the space of maps of the $(d-2)$-dimensional sphere $S^{d-2}$ to the border  $\partial\Omega$ of $\Omega$. Therefore, there must be some sort of Poincar\'e lemma playing a role here, implying that the curvature of a connection which is already a curvature should vanish. So, in that sense ${\cal J}$ would play the role of the flat connection in the approach proposed in \cite{afs,afs-review}. Note however that the integral form of the equations of motion \rf{stokes} does not require the introduction of a connection in loop space to obtain conserved quantities  for the theories we consider in this paper. 

The paper is organized as follows: in section \ref{sec:1+1} we implement our construction for integrable field theories in $1+1$  dimensions re-obtaining  well known results in that research area using the integral form of the equations of motion \rf{stokes}. In section \ref{sec:2+1} we discuss the cases of the Chern-Simons theory in the presence of a source as well as the Yang-Mills theories, both in $2+1$ dimensions. An important result of this section is the quantization of the charges in the case where the two dimensional space sub-manifold has no border. In section \ref{sec:3+1} we discuss the interesting case of  non-abelian gauge theories in $3+1$ dimensions, in the presence of matter currents. An important result here is an integral formulation of the Yang-Mills equations, and the construction of gauge invariant conserved charges. In the appendices \ref{sec:curves} and \ref{sec:surface} we give the proofs of the non-abelian Stokes theorems used in our constructions.

\section{The case of curves: theories in $1+1$ dimensions}
\label{sec:1+1}
\setcounter{equation}{0}

In a $1+1$ dimensional  space-time $M$ we establish a dynamical equation relating a field $g(x)$, element of a Lie group $G$,  to another field $C_{\mu}(x)$, a 1-form taking values in the Lie algebra $\mathcal{G}$ of $G$. The relation between those two fields is built as follows. Consider a path $\gamma$ in $M$,  parametrized by $\sigma$, and define a quantity $W$ through the differential equation
\begin{equation}
\lab{holo_w}
\frac{dW}{d\sigma}+C_{\mu}\frac{dx^{\mu}}{d\sigma}W=0
\end{equation}
where $x^{\mu}$, $\mu=0,1$, are the Cartesian coordinates in $M$ of the points of $\gamma$. 
Integration of \rf{holo_w} can be formally written as $W=P_{1}e^{-\int_{\gamma}d\sigma C_{\mu}\frac{dx^{\mu}}{d\sigma}}\cdot W_{R}$, where $P_{1}$ stands for the path-ordering, and $W_R$ is an integration constant corresponding to the value of $W$ at the initial point $x_R$ of the path $\gamma$.

Given any smooth path $\gamma$ in $M$, with initial and final points denoted by $x_R$ and $x_f$ respectively, we impose the following equation for the fields $g(x)$ and $C_{\mu}(x)$
\begin{equation}
\lab{stokes_1d}
g(x_{f})\cdot g(x_{R})^{-1}=P_{1}\;e^{-\int_{\gamma}d\sigma C_{\mu}(x)\frac{dx^{\mu}}{d\sigma}}
\end{equation}
with $g(x_{R})$ and $g(x_{f})$ corresponding to the values of $g(x)$ at the end points $x_{R}$ and $x_{f}$, and the r.h.s. of \rf{stokes_1d} is obtained by integrating \rf{holo_w} along the path $\gamma$, assuming that the integration constant is unit. Note that \rf{stokes_1d} has the form of \rf{stokes} since the border of $\gamma$ corresponds to its end points, i.e. $\partial\gamma=\{x_R, x_f\}$, and $g(x_{f})\cdot g(x_{R})^{-1}$ stands for the integration of  ${\cal F}$ on $\partial\gamma$, which in this case  would be a zero-form.

\begin{figure}[t]
  \begin{center}
    \includegraphics[width=0.4\textwidth]{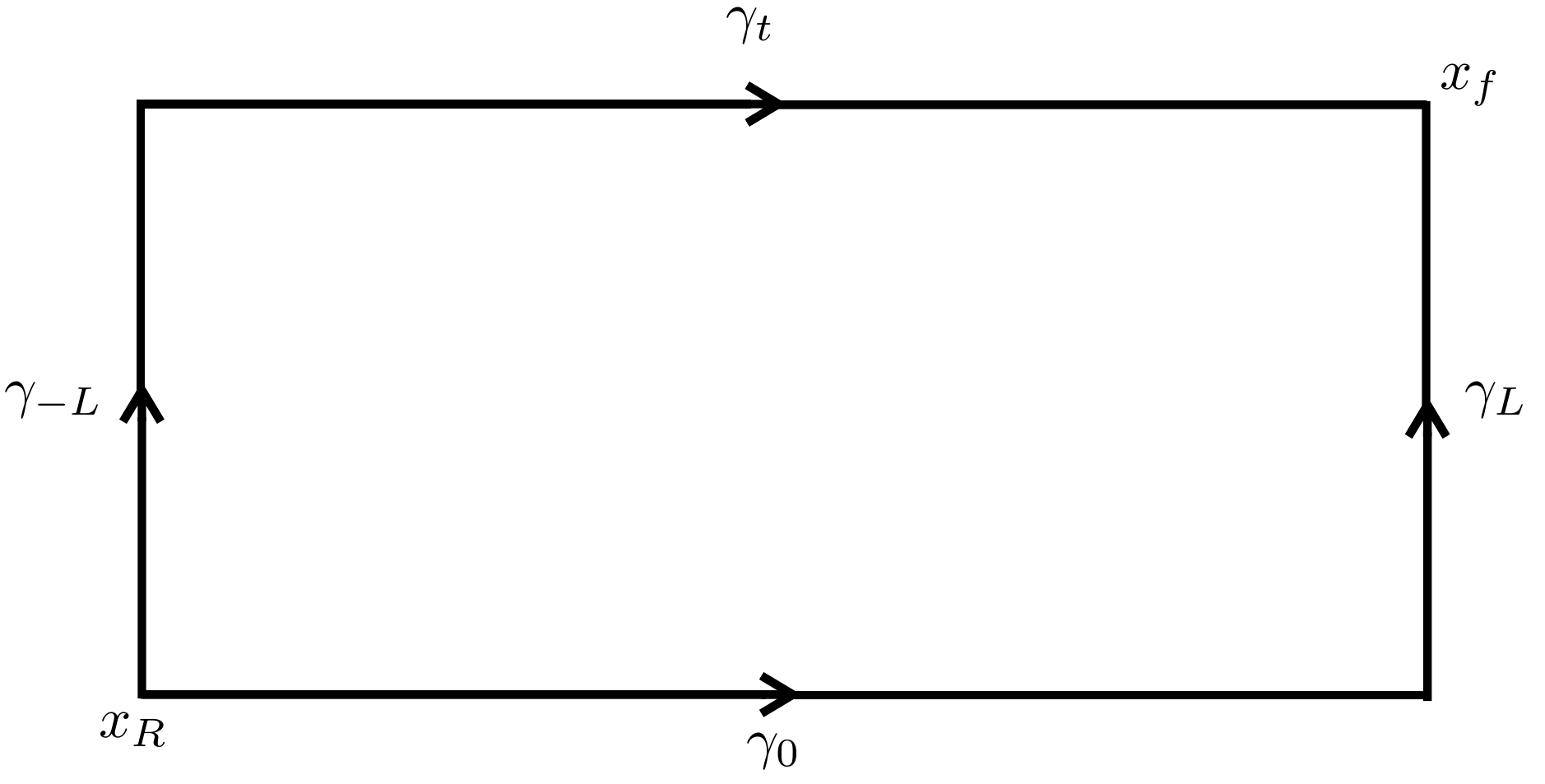}
  \end{center}
  \caption{The two paths $\gamma_{1}=\gamma_{L}\cdot \gamma_{{0}}$ and $\gamma_{2}=\gamma_{t}\cdot \gamma_{-L}$, connecting the points $x_R$ and $x_f$, which are used to construct the conserved charges as the eigenvalues of the operator \rf{charge1d}. The horizontal paths are parallel to the space axis and the vertical ones to the time axis. }
 \label{path_charge_1d} 
\end{figure}

The first important consequence of \rf{stokes_1d} is that the path ordered integral of $C_{\mu}(x)$ is independent of the path. Indeed, if $\gamma_1$ and $\gamma_2$ are two paths in $M$ with the same end points $x_{R}$ and $x_{f}$, then  \rf{stokes_1d} implies that $P_{1}\;e^{-\int_{\gamma_1}d\sigma C_{\mu}(x)\frac{dx^{\mu}}{d\sigma}}=P_{1}\;e^{-\int_{\gamma_2}d\sigma C_{\mu}(x)\frac{dx^{\mu}}{d\sigma}}$. That fact together with some appropriate boundary conditions is a conservation law as we now explain. Consider the space-time as being $M=\IR\times\IR$, and let $x^0\equiv t$ and $x^1\equiv x$ be the time and space coordinates respectively. We choose the coordinates of $x_R$ as being $(t,x)=(0,-L)$ and of $x_f$ as  $(t,x)=(t,L)$, with $L$ being a length scale which will be taken to infinity at the end of calculations. We choose two paths joining $x_R$ and $x_f$ as shown in Figure \ref{path_charge_1d}, i.e. the first path is $\gamma_{1}=\gamma_{L}\cdot \gamma_{{0}}$ and the second $\gamma_{2}=\gamma_{t}\cdot \gamma_{-L}$.  Note that $\gamma_t$ and $\gamma_{{0}}$ are paths at constant time at $t=t$ and $t=0$ respectively. On the other hand $\gamma_{L}$ and $\gamma_{-L}$ are paths at constant space at $x=L$ and $x=-L$ respectively. If one assumes that the time component of the one-form satisfies the boundary condition $C_0(t,-L)=C_0(t,L)$, for all values of $t$, then the path ordered integrals of $C_{\mu}$ along $\gamma_{L}$ and $\gamma_{-L}$ are the same, i.e. $P_{1}\;e^{-\int_{\gamma_{-L}}d\sigma C_{\mu}(x)\frac{dx^{\mu}}{d\sigma}}=P_{1}\;e^{-\int_{\gamma_L}d\sigma C_{\mu}(x)\frac{dx^{\mu}}{d\sigma}}\equiv U(t)$. Therefore, the equality of the path ordered integrals of $C_{\mu}$ along $\gamma_1$ and $\gamma_2$ leads to the iso-spectral evolution equation
\be
 P_{1}\;e^{-\int_{\gamma_{t}}d\sigma C_{\mu}(x)\frac{dx^{\mu}}{d\sigma}}= U(t)\,
 P_{1}\;e^{-\int_{\gamma_{0}}d\sigma C_{\mu}(x)\frac{dx^{\mu}}{d\sigma}}\,U(t)^{-1}
 \ee
Consequently the eigenvalues of the operator
\be
Q = P_{1}\;e^{-\int_{\gamma_{t}}d\sigma C_{\mu}(x)\frac{dx^{\mu}}{d\sigma}}= g(t,L)\,g^{-1}(t,-L)
\lab{charge1d} 
\ee
where in the last equality we have used \rf{stokes_1d}, are constant in time. Similarly, one can express those constants of motion as ${\rm Tr} Q^N$, for any integer $N$. When we take the limit $L\rightarrow \infty$, one observes that the conserved charges are determined by the asymptotic values of the field $g(x)$, which is a known result in soliton theory \cite{luizwojtek}.

Let us now consider the case where the space-time is of the form $M= {\cal S}\times \IR$, where the space submanifold ${\cal S}$ has no border, like for instance the circle ${\cal S}=S^1$.  It then follows from \rf{stokes_1d} that the path ordered integral of $C_{\mu}$ on the whole space ${\cal S}$ must be unity, i.e.
\be
P_{1}\;e^{-\int_{{\cal S}}d\sigma C_{\mu}(x)\frac{dx^{\mu}}{d\sigma}}=\one
\lab{charges1}
\ee
since the initial and final points are the same and so $g(x_R)=g(x_f)$. That is true for any value of time, and   consequently \rf{charges1} can be interpreted as a conservation law, where  the conserved charges are in fact trivial. However, depending upon the  theory under consideration, one gets (topological) quantization conditions for some quantities. A simple example would be that of an abelian pure imaginary connection, $C_{\mu}\equiv i\, J_{\mu}$, where \rf{charges1} leads to $\int_{{\cal S}} d\sigma J_{\mu}(x)\frac{dx^{\mu}}{d\sigma}=2\,\pi\,n$, with $n$ integer.

Note that \rf{stokes_1d} transforms covariantly under the gauge transformations 
\be
C_{\mu} \rightarrow h\,C_{\mu}\,h^{-1}-\partial_{\mu}h\,h^{-1}\qquad\qquad g\rightarrow h\, g
\ee
since \rf{holo_w} implies that $P_{1}\;e^{-\int_{\gamma}d\sigma C_{\mu}(x)\frac{dx^{\mu}}{d\sigma}}\rightarrow 
h(x_f)\,P_{1}\;e^{-\int_{\gamma}d\sigma C_{\mu}(x)\frac{dx^{\mu}}{d\sigma}}\,h(x_R)^{-1}$, with $x_R$ and $x_f$ being the end points of $\gamma$. Therefore, the conserved charges given by the eigenvalues of \rf{charge1d} are invariant under those gauge transformations satisfying $h(t,-L)=h(t,L)$.

If $\gamma$ is taken to be a path infinitesimally short, so that its end points approach each other, then $g$ at $x_{f}$ can be written as an approximation of the value it has at $x_{R}$ by using a Taylor expansion: $g(x_{f})=g(x_{R})+\partial_{\mu}g(x_{R})\delta x^{\mu}$, with $\delta x^{\mu}$ being the infinitesimal displacement between $x_f$ and $x_R$. Therefore, the l.h.s. of \rf{stokes_1d}, up to first order in $\delta x^{\mu}$, becomes $g(x_{f})g(x_{R})^{-1} \sim \one + \partial_{\mu}g(x_{R})\, g(x_{R})^{-1}\, \delta x^{\mu}$. In addition,  the path-ordering effects in the integration of $C_{\mu}$ are of higher order in $\delta x^{\mu}$, and therefore the r.h.s of \rf{stokes_1d}, up to first order, becomes simply ${\one} -C_{\mu}\delta x^{\mu}$. Since that is valid for any infinitesimal path located anywhere in the space-time $M$, we get that \rf{stokes_1d}  implies the  following differential equation for the fields $g(x)$ and $C_{\mu}(x)$ 
\begin{equation}
C_{\mu}(x)=-\pa_{\mu}g(x)\;g^{-1}(x)
\lab{puregage}
\end{equation}
Therefore $C_{\mu}$ is of the form of a pure gauge field, and consequently its curvature vanishes, i.e. 
\be
\pa_{\mu}C_{\nu}-\pa_{\nu}C_{\mu}+[C_{\mu},C_{\nu}]=0
\lab{lax}
\ee
The relation \rf{lax} is the so-called Lax-Zakharov-Shabat equation \cite{lax} or the zero curvature condition, which is the basic structure used in the development of exact methods in soliton theory and two dimensional integrable field theories. The equation \rf{stokes_1d} is therefore an integral formulation of the Lax-Zakharov-Shabat equation. One can in fact obtain one from the other. However, there are some subtleties in the integral formulation, since it works with the two fields $C_{\mu}$ and $g(x)$ and a relation between them, namely    \rf{stokes_1d}. That approach leads in a quite natural way,  as shown in \rf{charge1d}, to the fact that  the conserved  charges come from boundary terms. Such  result is known for a large class of soliton theories \cite{luizwojtek}, but it is not so certain that it holds for integrable field theories not possessing solitons. It would be interesting to investigate that issue further. In addition, the integral formulation leads  to the triviality of the charges, or its topological quantization, in the case where the space sub-manifold has no border.

\section{The case of surfaces: theories in $2+1$ dimensions} 
\label{sec:2+1}
\setcounter{equation}{0}

In the case of theories defined on a $(2+1)$-dimensional space-time $M$ the basic ingredient of our construction is the so-called non-abelian Stokes theorem  for a one-form connection $C_{\mu}$. Let $\Sigma$ be a two dimensional smooth surface on $M$, and let $\partial \Sigma$ be its border, i.e. a closed curve on $M$.   The theorem states that the path-ordered integral of $C_{\mu}$ around $\partial \Sigma$ is equal to the surface ordered integral on $\Sigma$, of the curvature of $C_{\mu}$, i.e. 
\be
P_{1}e^{-\oint_{\pa \Sigma}d\sigma\;C_{\mu}\frac{dx^{\mu}}{d\sigma}}\cdot W_{R}=W_{R}\cdot P_{2}\;e^{\int_{\Sigma}d\tau\;d\sigma\;W^{-1}\,G_{\mu\nu}\, W\,\frac{dx^{\mu}}{d\sigma}\frac{dx^{\nu}}{d\tau}}
\lab{usual_stokes}
\ee
where
\be
G_{\mu\nu}=\partial_{\mu}C_{\nu}-\partial_{\nu}C_{\mu}+\sbr{C_{\mu}}{C_{\nu}}
\lab{gdef}
\ee
A proof of \rf{usual_stokes} is given in the appendix \ref{sec:curves}, but its meaning is the following.  One chooses a reference point $x_R$ on the border  of $\Sigma$ and scan it with closed loops starting and ending at $x_R$. The loops are labelled by $\tau$ such that $\tau=0$ corresponds to the infinitesimal loop around $x_R$, and $\tau=2\pi$ corresponds to the border $\partial\Sigma$. Each closed loop is parametrized by $\sigma$ such that $\sigma=0$ and $\sigma=2\pi$ corresponds to $x_R$. The l.h.s. of \rf{usual_stokes} is obtained by integrating the differential equation \rf{holo_w} along $\partial\Sigma$, and $W_R$ is the integration constant corresponding to the value of $W$ at $x_R$. The meaning of $P_1$ is that such integration has to be path ordered. As shown in the appendix \ref{sec:curves}, the r.h.s. of \rf{usual_stokes} is obtained by  integrating on  $\Sigma$ the differential equation 
\begin{equation}
\lab{holo_v}
\frac{dV}{d\tau}-V\mathcal{J}=0
\end{equation}
with
\be
\mathcal{J}\equiv \int_{0}^{2\pi}d\sigma\;W^{-1}G_{\mu\nu}W\frac{dx^{\mu}}{d\sigma}\frac{dx^{\nu}}{d\tau}
\lab{jcaldef}
\ee
and the meaning of $P_2$ is that such integration has to be surface ordered according to the scanning of $\Sigma$ with loops as explained above. 
Again $W_R$ is the integration constant and corresponds to the value of $V$ on the infinitesimal loop around $x_R$. That the two integration constants have to be the same can be understood by shrinking $\Sigma$ to the reference point $x_R$. 

We now show how to use the non-abelian Stokes theorem \rf{usual_stokes} to define an integral formulation of the Chern-Simons and Yang-Mills theories, both in the presence of sources, and on a space-time of  $2+1$ dimensions. We then show how to use  \rf{usual_stokes} to   construct conserved charges for those theories.

\subsection{Integral formulation of Chern-Simons theory with matter source}
\label{sec:cs}

Consider a theory on a $(2+1)$-dimensional space-time $M$ for a vector field $A_{\mu}$ and a current $J_{\mu}$, with $\mu=0,1,2$, and let its classical equations of motion be defined as follows. On any two dimensional smooth surface $\Sigma$ on $M$, with border $\partial\Sigma$, the fields must satisfy the integral equations
\begin{equation}
\lab{cs_int}
P_{1}e^{-ie\oint_{\pa \Sigma}d\sigma\;A_{\mu}\frac{dx^{\mu}}{d\sigma}}= P_{2}\;e^{\frac{ie}{\kappa}\int_{\Sigma}d\tau\;d\sigma\;W^{-1}\widetilde{J}_{\mu\nu}W\frac{dx^{\mu}}{d\sigma}\frac{dx^{\nu}}{d\tau}}.
\end{equation}
where $\widetilde{J}_{\mu\nu}$ stands for the Hodge dual of the matter current i.e., $\widetilde{J}_{\mu\nu}\equiv \epsilon_{\mu\nu\rho}J^{\rho}$, and where $e$ and $\kappa$ are coupling constants of the theory.  The meaning of the path ordered ($P_1$) and surface ordered ($P_2$) integrals in \rf{cs_int} is the same as those in  \rf{usual_stokes}, i.e.  the l.h.s. of \rf{cs_int} is obtained by integrating \rf{holo_w} with $C_{\mu}=i\,e\,A_{\mu}$, and its r.h.s. by integrating \rf{holo_v} with $G_{\mu\nu}=\frac{i\,e}{\kappa}\,\widetilde{J}_{\mu\nu}$. 

Since \rf{cs_int} is valid on any $\Sigma$, then it has to hold true when $\Sigma$ is taken to be an infinitesimal surface. It then follows that \rf{cs_int} implies local differential equations for the fields as we now explain.  Indeed, take $\Sigma$ to be a planar surface of rectangular shape on the plane defined by two axis of the Cartesian coordinates, let us say $x^{\mu}$ and $x^{\nu}$, with $\mu$ and $\nu$ fixed.   The border $\partial\Sigma$ is then a rectangle of infinitesimal sides $\delta x^{\mu}$ and $\delta x^{\nu}$. We evaluate both sides of \rf{cs_int} by Taylor expanding the integrands around one given corner of the rectangle, and keeping things at the lowest non-trivial order. One can check that  the l.h.s. of \rf{cs_int} becomes $\one+i\,e\,F_{\mu\nu}\delta x^{\mu}\delta x^{\nu}$, with no sum in $\mu$ and $\nu$, and where 
\be
F_{\mu\nu}=\partial_{\mu}A_{\nu}-\partial_{\nu}A_{\mu}+i\,e\,\sbr{A_{\mu}}{A_{\nu}}
\ee
The r.h.s. of \rf{cs_int} in lowest order is given by $\one+\frac{i\,e}{\kappa}\,\widetilde{J}_{\mu\nu}\delta x^{\mu}\delta x^{\nu}$ (no sum in $\mu$ and $\nu$). Therefore, for an infinitesimal surface, \rf{cs_int} implies the local differential equations for $A_{\mu}$  
\be
F_{\mu\nu}= \frac{1}{\kappa}\,\widetilde{J}_{\mu\nu}=\frac{1}{\kappa}\,\epsilon_{\mu\nu\rho}J^{\rho}
\lab{cs_diff}
\ee
which are the equations of motion of the Chern-Simons theory in the presence of an external current $J_{\mu}$. 

On the other hand one observes that if one takes $C_{\mu}$ in \rf{usual_stokes} as $C_{\mu}=i\,e\,A_{\mu}$, and therefore $G_{\mu\nu}=i\,e\,F_{\mu\nu}$, and uses  \rf{cs_diff}, then one obtains \rf{cs_int}. In other words, \rf{cs_int} is a direct consequence of the non-abelian Stokes theorem \rf{usual_stokes} and the Chern-Simons equations of motion \rf{cs_diff}. Since \rf{cs_int} implies \rf{cs_diff}, we see that \rf{cs_int} is indeed an integral formulation of the Chern-Simons theory.  

Note that in obtaining \rf{cs_int} from the non-abelian Stokes theorem \rf{usual_stokes} we have dropped the integration constant $W_R$. That has to do with the covariance of \rf{cs_int} under gauge transformations, as we now explain. The Chern-Simons equation \rf{cs_diff} transforms covariantly under the gauge transformations $A_{\mu}\rightarrow g\,A_{\mu}\,g^{-1}+\frac{i}{e}\,\partial_{\mu}g\,g^{-1}$, since $F_{\mu\nu}\rightarrow g\, F_{\mu\nu}\,g^{-1}$, and $\widetilde{J}_{\mu\nu}\rightarrow g\, \widetilde{J}_{\mu\nu}\,g^{-1}$. From \rf{holo_w} we have that under a gauge transformation $W\rightarrow g_f\, W\,g_i^{-1}$, where $g_i$ and $g_f$ are the values of $g$ at the initial and final points of the curve where $W$ is defined. Consequently,  on a closed curve one has that $W^c\rightarrow g_R\, W^c\,g_R^{-1}$, where $g_R$ is the value of $g$ at the reference point $x_R$, where the curve starts and ends.  In addition, $\mathcal{J}$  defined in \rf{holo_w}, with $G_{\mu\nu}$ replaced by $\frac{i\,e}{\kappa}\,\widetilde{J}_{\mu\nu}$, transforms as $\mathcal{J}\rightarrow g_R\,\mathcal{J}\,g_R^{-1}$, and so from \rf{holo_v} we have that $V\rightarrow g_R\,V\,g_R^{-1}$. However, if $W^c_1$ and $V_1$ are solutions of \rf{holo_w} and \rf{holo_v} respectively, so are $W^c_2=W^c_1\, k$ and $V_2=h\,V_1$, with $k$ and $h$ constant group elements. Under a gauge transformation one would then have $W^c_i\rightarrow g_R\, W^c_i\,g_R^{-1}$, and $V_i\rightarrow g_R\,V_i\,g_R^{-1}$, with $i=1,2$. But since $k$ and $h$ are arbitrary group elements one should not expect them to depend upon $A_{\mu}$, and so be insensitive to its gauge transformations. Therefore, one could as well conclude that $W^c_2\rightarrow g_R\, W^c_1\,g_R^{-1}\, k$, and $V_2\rightarrow h\, g_R\,V_1\,g_R^{-1}$. The only way to establish a compatibility is to assume that $k$ and $h$ should belong to the center of the gauge group $G$, since $g_R$ can be any element of $G$.  Since the integration constants $W_R$ in \rf{usual_stokes} have the same status in this discussion, as $k$ and $h$, we have to take  them to lie in the center of $G$ to have the gauge covariance of the integral Chern-Simons equation \rf{cs_int}. However, when that is done they drop out from \rf{cs_int} since they commute with the path and surface ordered integrals. So, when integrating \rf{holo_w} and \rf{holo_v} to construct the l.h.s. and r.h.s. respectively  of \rf{cs_int} one should keep in mind that those operators can carry an integration constant lying in the center of $G$ without destroying the gauge covariance of \rf{cs_int}. That fact may be important in some applications.

Note that both sides of \rf{cs_int} depend upon the choice of the reference point $x_R$ and also on the choice of the scanning of $\Sigma$ with loops. However, when one changes the scanning and the reference point, the non-abelian Stokes theorem \rf{usual_stokes} guarantees that both sides of \rf{cs_int} change in a way that they remain equal to each other. In that sense one can say that \rf{cs_int} transforms ``covariantly'' under the change of scanning and reference point. 
In fact, even though we have   defined the equation \rf{cs_int} on any surface $\Sigma$ in $M$, it is formally defined on the loop space $\mathcal{L}\Sigma = \{ \gamma: \mathcal{S}^{1} \rightarrow \Sigma \; \vert \; \text{north pole} \rightarrow x_{R} \in \partial \Sigma \}$, consisting of maps from  the circle $\mathcal{S}^{1}$ into $\Sigma$, such that the north pole of $\mathcal{S}^{1}$  is mapped into $x_R$. The images of such maps are closed loops in $\Sigma$, starting and ending at $x_R$. Since $\Sigma$ is scanned by a collection of such loops, and the loops are points in  $\mathcal{L}\Sigma$, one can see $\Sigma$ as a path in $\mathcal{L}\Sigma$. Therefore, a change in the scanning of $\Sigma$ corresponds to a change of path in $\mathcal{L}\Sigma$ representing the same physical $\Sigma$. 

Despite the fact that \rf{cs_int} is defined on loop space, it leads to very physical consequences, like conservation laws as we now explain. Consider the case where the surface $\Sigma$ is a closed surface $\Sigma_c$, i.e. with no border. Then, the l.h.s. of \rf{cs_int} is trivial and we are lead to 
\begin{equation}
\lab{charge_2d}
P_{2}\;e^{\frac{ie}{\kappa}\oint_{\Sigma_{\text{c}}}d\tau\;d\sigma\;W^{-1}\widetilde{J}_{\mu\nu}W\frac{dx^{\mu}}{d\sigma}\frac{dx^{\nu}}{d\tau}}=\one 
\end{equation}
Being a closed surface, $\Sigma_c$ corresponds to a closed path in the loop space  $\mathcal{L}\Sigma_c$, starting and ending at the reference point $x_R$. Consider now a point on that path corresponding to a loop  
$\gamma$ in $\Sigma_c$. It divide the path in two parts corresponding to two surfaces, i.e. we have $\Sigma_c=\Sigma_1+\Sigma_2$. From the ordering defined by \rf{holo_v} we have that
\be
P_{2}\;e^{\frac{ie}{\kappa}\int_{\Sigma_1}d\tau\;d\sigma\;W^{-1}\widetilde{J}_{\mu\nu}W\frac{dx^{\mu}}{d\sigma}\frac{dx^{\nu}}{d\tau}}\;P_{2}\;e^{\frac{ie}{\kappa}\int_{\Sigma_2}d\tau\;d\sigma\;W^{-1}\widetilde{J}_{\mu\nu}W\frac{dx^{\mu}}{d\sigma}\frac{dx^{\nu}}{d\tau}}=\one
\lab{quasiindependent}
\ee
By reverting the sense of the integration along the path,  one obtains  the inverse operator when integrating \rf{holo_v}. Then, $\Sigma_1$ and $\Sigma_2^{-1}$ are two surfaces corresponding to two paths starting and ending at the same points, namely the reference point $x_R$ and the loop $\gamma$, which is their common border. Therefore, \rf{quasiindependent} implies that the integration along two paths with the same end points gives the same operator.  Since that is valid for any closed surface $\Sigma_c$ and any partition of it into two surfaces, we conclude that the quantity 
$P_{2}\;e^{\frac{ie}{\kappa}\int_{\Sigma}d\tau\;d\sigma\;W^{-1}\widetilde{J}_{\mu\nu}W\frac{dx^{\mu}}{d\sigma}\frac{dx^{\nu}}{d\tau}}$ is independent of the surface $\Sigma$. It depends only on the initial and final points of the path corresponding to $\Sigma$ in loop space, i.e. the border $\partial \Sigma$ and the reference point $x_R$ on it. Note that such independency corresponds to the change of the reparameterization  (scanning) of the surface, as well as to the change of the surface itself, but keeping the border and reference point fixed. 
Such surface independency leads to conservation laws as we now explain.

First we consider the topology of space-time to be $M=\mathcal{S}\times \IR$, with time being the real line $\IR$, and the space being the closed two dimensional surface with no boundary $\mathcal{S}$. A simple case is when $\mathcal{S}$ is the two-sphere, i.e. $\mathcal{S}=S^2$. Then if we evaluate \rf{charge_2d} in space (\emph{i.e.}, $\Sigma_{\text{c}}=\mathcal{S}$) we get that the quantity $Q_{\mathcal{S}}\equiv P_{2}\;e^{\frac{ie}{\kappa}\oint_{\mathcal{S}}d\tau\;d\sigma\;W^{-1}\widetilde{J}_{\mu\nu}W\frac{dx^{\mu}}{d\sigma}\frac{dx^{\nu}}{d\tau}}=\one$, is  conserved in time and equal to unity. That is an interesting relation since it may imply that the net charge on the whole space vanishes, or then that the charge must satisfy some quantization condition. 

For simplicity, let us consider the case of an  abelian theory, with gauge group $G=U(1)$, where the path and surface orderings are irrelevant. In such a case one has  $Q_{\mathcal{S}}=\;e^{\frac{ie}{\kappa}q}=1$, where $q$ is the total charge in space, i.e. $q=\oint_{\mathcal{S}}d\tau\;d\sigma\;\widetilde{J}_{\mu\nu}\frac{dx^{\mu}}{d\sigma}\frac{dx^{\nu}}{d\tau}$. Those equations  establishes a quantization condition involving the total amount of charge in space and the Chern-Simons coupling constants, i.e. 
\be
\frac{q\,e}{\kappa}= 2\,\pi\, n \qquad\qquad \qquad \mbox{\rm with $n$ integer}
\lab{cs_quant}
\ee
Such result has two important consequences for Chern-Simons theory on the space-time $M=\mathcal{S}\times \IR$. The equations of motion \rf{cs_diff} imply that the time component $J^0$ of the current, namely the charge density $\rho$, is proportional to the space components of the field tensor, which is the pseudo-scalar magnetic field $B$, i.e. $\rho=\kappa\, B$. So, the effect of the Chern-Simons equation of motion is to attach magnetic flux to the electric charge \cite{dunne}. For a point particle we have $\rho=e \,\delta({\vec x}-{\vec x}_1)$, with ${\vec x}_1$ being the position vector of the particle. Therefore, the magnetic flux associated to such a particle is $\Phi=\frac{e}{\kappa}$, and \rf{cs_quant} implies it is quantized as $\Phi=\frac{2\,\pi\,n}{q}$.
The second  consequence of \rf{charge_2d} and the topology $M=\mathcal{S}\times \IR$,   is that the phase gained by a non-relativistic particle that moves around another, due to a Aharonov-Bohm type interaction,  is no longer dependent on the Chern-Simons coupling constant $\kappa$. For $N$ such particles the charge density reads $\rho=e\sum^{N}_{a}\delta (\vec{x}-\vec{x}_{a})$ and the attached magnetic field $B=\frac{e}{\kappa}\sum^{N}_{a}\delta (\vec{x}-\vec{x}_{a})$. After a double interchange of two particles, their phase exchange is given by $\Delta \theta = \frac{e^{2}}{4\pi \kappa}$ \cite{dunne}. Due to the  quantization condition \rf{cs_quant} and using that $q=Ne$ we get $\Delta \theta = \frac{n}{2N} $, which is a rational number, and not any number as would be the case if \rf{charge_2d}, and so \rf{cs_quant}, was not used.

\begin{figure}[t]
  \begin{center}
    \includegraphics[width=0.7\textwidth]{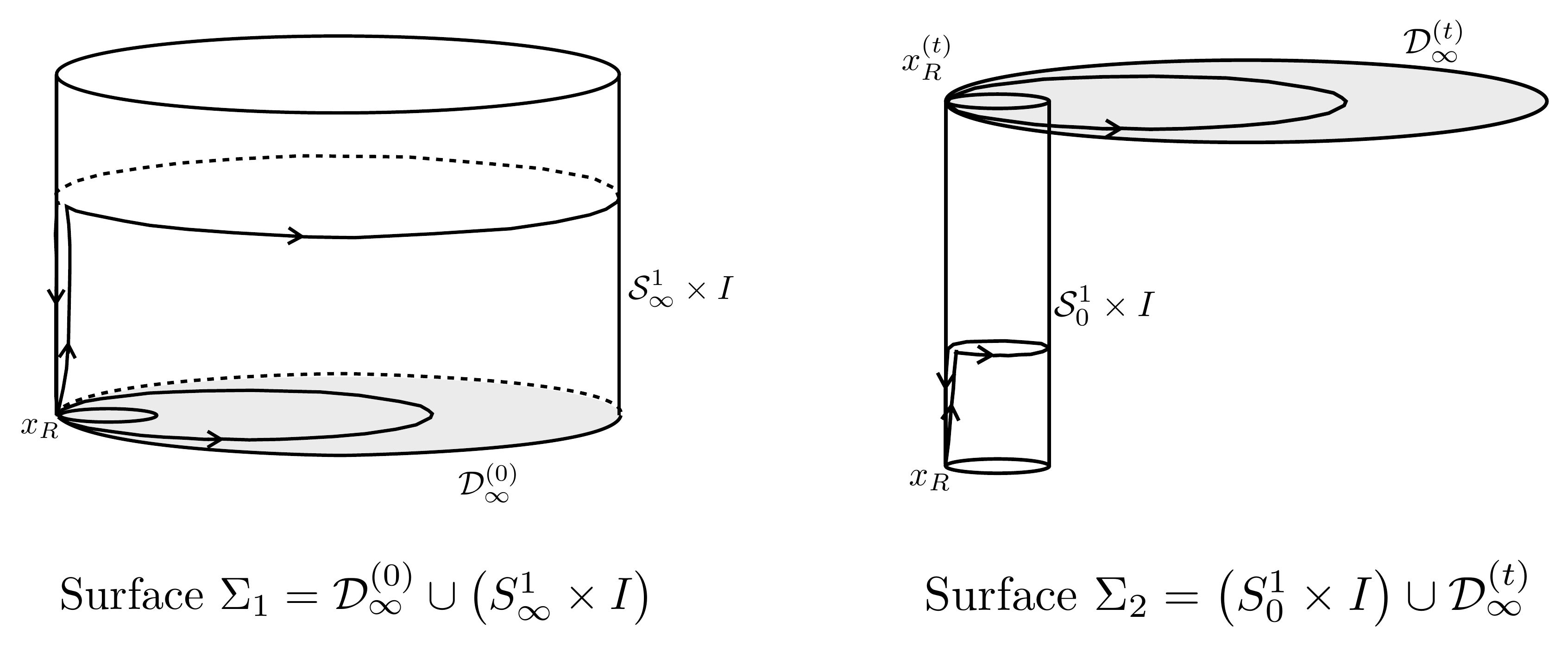}
  \end{center}
  \caption{The surfaces $\Sigma_1$ and $\Sigma_2$, with the same border $S^{1,(t)}_{\infty}$, and reference point $x_R$, used in the construction of conserved charges.}
 \label{fig:surface_scan} 
\end{figure}

Let us now consider the case where space-time  is $\IR^{3}$, and discuss how conserved charges can be constructed using \rf{charge_2d}. As we have seen, as a consequence of  \rf{charge_2d}, the quantity 
\be
V\( \Sigma\)\equiv P_{2}\;e^{\frac{ie}{\kappa}\int_{\Sigma}d\tau\;d\sigma\;W^{-1}\widetilde{J}_{\mu\nu}W\frac{dx^{\mu}}{d\sigma}\frac{dx^{\nu}}{d\tau}}
\lab{vdefindepend}
\ee
 is independent of the surface $\Sigma$, as long as its border $\partial\Sigma$ is kept fixed and also the reference point $x_R$ on it. In addition it is independent of the scanning (parameterization) of $\Sigma$ with loops. We shall consider two surfaces, $\Sigma_1$ and $\Sigma_2$, with the same borders as follows. The first one, shown in figure  \ref{fig:surface_scan}, is made of two parts. The first part is a disk ${\cal D}_{\infty}^{(0)}$ on the plane $x^1\,x^2$ at time $x^0=0$, and of a radius which will be taken to be infinite. The second part is a cylinder $S^1_{\infty}\times I$, where $I$ is a segment of the $x^0$-axis going from $x^0=0$ to $x^0=t$,  and $S^1_{\infty}$ is a circle of infinite radius parallel to the plane $x^1\,x^2$. We take the reference point $x_R$ to be on the border of the disk ${\cal D}_{\infty}^{(0)}$, as shown in \ref{fig:surface_scan}, and scan $\Sigma_1={\cal D}_{\infty}^{(0)}\cup \(S^1_{\infty}\times I\)$, with loops starting and ending at $x_R$, labeled by $\tau$, such that for $\tau\in \left[ 0 ,\pi\right]$ we scan ${\cal D}_{\infty}^{(0)}$, with $\tau=0$ corresponding to the infinitesimal loop around $x_R$, and $\tau=\pi$ corresponding to $S^1_{\infty}$, the border of ${\cal D}_{\infty}^{(0)}$. For $\tau\in \left[ \pi ,2\,\pi\right]$ we scan $S^1_{\infty}\times I$ with loops, which start at $x_R$, go up in the $x^0$ direction upon to  $x^0=t^{\prime}\in I$, go round $S^1_{\infty}$, and come down to $x_R$ again. By varying  $t^{\prime}$ within $I$ we scan the cylinder $S^1_{\infty}\times I$. Following the notation of \rf{vdefindepend}, and the ordering defined by \rf{holo_v}, one gets
 \be
 V\(\Sigma_1\)=V\({\cal D}_{\infty}^{(0)}\)\,V\(S^1_{\infty}\times I\)
 \lab{vsigma1}
 \ee
The second surface $\Sigma_2$, as shown in figure \ref{fig:surface_scan}, is also made of two parts. The first part is a cylinder $S^1_0\times I$, with $I$ being the same time interval as above, and $S^1_0$ a circle parallel to the $x^1\,x^2$ plane with infinitesimal radius. The reference point $x_R$ is on the border of the base of such cylinder at $x^0=0$. The second part is a disk ${\cal D}_{\infty}^{(t)}$ on the plane $x^1\,x^2$ at time $x^0=t$, and of a radius which will be taken to be infinite. The surface $\Sigma_2=\(S^1_0\times I\)\cup {\cal D}_{\infty}^{(t)}$ is scanned with loops starting and ending at $x_R$, labelled by $\tau$, such that for $\tau\in \left[ 0, \pi\right]$, we scan $S^1_0\times I$ with loops, which start at $x_R$, go up in the $x^0$ direction upon to  $x^0=t^{\prime}\in I$, go round $S^1_{0}$, and come down to $x_R$ again. By varying  $t^{\prime}$ within $I$ we scan the cylinder $S^1_{0}\times I$. For $\tau\in \left[ \pi, 2\,\pi\right]$ we scan ${\cal D}_{\infty}^{(t)}$ with loops which start at $x_R$, go up to $x^0=t$, go round a closed loop on ${\cal D}_{\infty}^{(t)}$, and come down to $x_R$ again. By keeping the two legs going up and down fixed and varying the closed loops  on ${\cal D}_{\infty}^{(t)}$, we scan it entirely. Again, following the notation of \rf{vdefindepend}, and the ordering defined by \rf{holo_v}, one gets
 \be
 V\(\Sigma_2\)=\,V\(S^1_{0}\times I\)\,V\({\cal D}_{\infty}^{(t)}\)
 \lab{vsigma2}
 \ee
Since $\Sigma_1$ and $\Sigma_2$ have the same reference point and the same border, namely $S^1_{\infty}$ at $x^0=t$, we have from the surface independency of \rf{vdefindepend} that $V\(\Sigma_1\)=V\(\Sigma_2\)$. We now impose the following boundary condition on our system
\be
{\tilde J}_{12}=J_0 \sim \frac{1}{r^{2+\delta}} \, T\({\hat r}\)\qquad\qquad {\rm for} \quad r\rightarrow \infty
\ee
with $\delta>0$, $r^2=\(x^1\)^2 +\(x^2\)^2$,   and $T\({\hat r}\)$ being an element of the Lie algebra of $G$, depending on the spatial direction defined by ${\hat r}=\frac{{\vec r}}{r}$. That condition implies that the quantity $\mathcal{J}\equiv \frac{i\,e}{\kappa}\,\int_{0}^{2\pi}d\sigma\;W^{-1}{\tilde J}_{\mu\nu}W\frac{dx^{\mu}}{d\sigma}\frac{dx^{\nu}}{d\tau}$, vanishes on loops at spatial infinity, and therefore  from \rf{holo_v} one gets that $V\(S^1_{\infty}\times I\)=\one$. In addition, since the circle $S^1_{0}$ has vanishing radius we also get that $V\(S^1_{0}\times I\)=\one$. Therefore, from the equality of \rf{vsigma1} and \rf{vsigma2} we get that $V\({\cal D}_{\infty}^{(t)}\)=V\({\cal D}_{\infty}^{(0)}\)$. Those two operators are calculated using the same reference point $x_R$, which lies at the border of ${\cal D}_{\infty}^{(0)}$. Consider now a reference point $x_R^{(t)}$, which have the same space coordinates as $x_R$, but at a time $x^0=t$, i.e. it lies on the border of  ${\cal D}_{\infty}^{(t)}$, just above $x_R$ (see figure \ref{fig:surface_scan}). By changing the reference point the quantity $\mathcal{J}$ changes as $\mathcal{J}\rightarrow W^{-1}(x_R^{(t)},x_R)\,\mathcal{J}\,W(x_R^{(t)},x_R)$, where $W(x_R^{(t)},x_R)$ is obtained by integrating $\rf{holo_w}$ on the path joining  $x_R$ to $x_R^{(t)}$. Therefore, if one  now integrates \rf{holo_v} on ${\cal D}_{\infty}^{(t)}$ with this new reference point, one gets that $V_{x_R}({\cal D}_{\infty}^{(t)})=W^{-1}(x_R^{(t)},x_R)\,V_{x_R^{(t)}}({\cal D}_{\infty}^{(t)})\,W(x_R^{(t)},x_R)$. Therefore, one gets that
\be
V_{x_R^{(t)}}\({\cal D}_{\infty}^{(t)}\)=W\(x_R^{(t)},x_R\)\,V_{x_R}\({\cal D}_{\infty}^{(0)}\) \,W^{-1}\(x_R^{(t)},x_R\)
\lab{isospectralv}
\ee
where the subindices indicate which reference point is being used in the integration of \rf{holo_v}. Note that in this way $V_{x_R^{(t)}}\({\cal D}_{\infty}^{(t)}\)$ and $V_{x_R}\({\cal D}_{\infty}^{(0)}\) $ correspond to surface ordered integrals over the entire space at times $x^0=t$ and $x^0=0$ respectively, and with reference points at spatial infinity (border of the infinite disks) and at the same times. Consequently \rf{isospectralv} constitute an iso-spectral time evolution for the operator 
\be
V_{x_R^{(t)}}\({\cal D}_{\infty}^{(t)}\)=P_{2}\;e^{\frac{ie}{\kappa}\int_{{\cal D}_{\infty}^{(t)}}d\tau\;d\sigma\;W^{-1}\widetilde{J}_{\mu\nu}W\frac{dx^{\mu}}{d\sigma}\frac{dx^{\nu}}{d\tau}}=
P_{1}e^{-ie\oint_{S^1_{\infty}}d\sigma\;A_{\mu}\frac{dx^{\mu}}{d\sigma}}
\lab{eigenconstant_cs}
\ee
where in the last equality we have used the Chern-Simons integral equation \rf{cs_int}, and where the spatial circle with infinite radius $S^1_{\infty}$ stands for the border of ${\cal D}_{\infty}^{(t)}$.  
Therefore, its eigenvalues, or equivalently ${\rm Tr}\left[V_{x_R^{(t)}}\({\cal D}_{\infty}^{(t)}\)\right]^N$, are constant in time. Those are the conserved quantities for the Chern-Simons theory. Note that such conserved quantities are gauge invariant, since under a gauge transformation we have that $V_{x_R^{(t)}}\({\cal D}_{\infty}^{(t)}\)\rightarrow  g\(x_R^{(t)}\)\, V_{x_R^{(t)}}\({\cal D}_{\infty}^{(t)}\)\,g\(x_R^{(t)}\)^{-1}$, where $g\(x_R^{(t)}\)$ is the element of the gauge group, performing the gauge transformation, evaluated at the reference point $x_R^{(t)}$. The conserved quantities are also independent of the way we scan ${\cal D}_{\infty}^{(t)}$, since we have already shown above that the operators of the type \rf{vdefindepend} are scanning independent. In fact, it was that property that lead to the conservation laws.  In addition, the conserved quantities are independent of the choice of the reference point on the border of ${\cal D}_{\infty}^{(t)}$. That has to with the fact that by changing the reference point, $V_{x_R^{(t)}}\({\cal D}_{\infty}^{(t)}\)$ changes by conjugation by an element $W$ obtained by integrating \rf{holo_w} along a path on the border joining the two reference points. Consequently, its eigenvalues are unchanged.

\subsection{Integral formulation of Yang-Mills theory with matter source}
\label{sec:ym2p1}

We now consider another theory on a $(2+1)$-dimensional space-time $M$, with the same field content, i.e.  a vector field $A_{\mu}$ and a current $J_{\mu}$, with $\mu=0,1,2$, and with  its classical equations of motion being defined as follows. On any two dimensional smooth surface $\Sigma$ on $M$, with border $\partial\Sigma$, the fields must satisfy the integral equations
\begin{equation}
\lab{ym_int}
P_{1}e^{-ie\oint_{\pa \Sigma}d\sigma\;\left(A_{\mu}+\beta\, {\widetilde F}_{\mu}\right)\,\frac{dx^{\mu}}{d\sigma}}= P_{2}\;e^{ie\,\int_{\Sigma}d\tau\;d\sigma\;W^{-1}\left(F_{\mu\nu}-\beta\,\widetilde{J}_{\mu\nu}+ie\,\beta^2\sbr{{\widetilde F}_{\mu}}{{\widetilde F}_{\nu}}\right)\,W\frac{dx^{\mu}}{d\sigma}\frac{dx^{\nu}}{d\tau}}.
\end{equation}
where $e$ is the coupling constant of the theory, $\beta$ is a free parameter, $\widetilde{J}_{\mu\nu}$ is the Hodge dual of the matter current i.e., $\widetilde{J}_{\mu\nu}\equiv \varepsilon_{\mu\nu\rho}J^{\rho}$,  ${\widetilde F}_{\mu}$ is the Hodge dual of the curvature of the connection, i.e. ${\widetilde F}_{\mu}=\frac{1}{2}\varepsilon_{\mu\nu\rho}F^{\nu\rho}$, and $F_{\mu\nu}=\pa_{\mu}A_{\nu}-\pa_{\nu}A_{\mu}+ie\,\sbr{A_{\mu}}{A_{\nu}}$.  The meaning of the path ordered ($P_1$) and surface ordered ($P_2$) integrals in \rf{ym_int} is the same as those in  \rf{usual_stokes}, i.e.  the l.h.s. of \rf{cs_int} is obtained by integrating \rf{holo_w} with 
\be
C_{\mu}=i\,e\,\(A_{\mu}+\beta\, {\widetilde F}_{\mu}\)
\lab{cdefym}
\ee
and its r.h.s. by integrating \rf{holo_v} with 
\be
G_{\mu\nu}=i\,e\(F_{\mu\nu}-\beta\,\widetilde{J}_{\mu\nu}+ie\,\beta^2\sbr{{\widetilde F}_{\mu}}{{\widetilde F}_{\nu}}\) 
\lab{gdefym}
\ee

In order to obtain the corresponding local equations of motion  we consider the integral equation  \rf{ym_int} on an infinitesimal surface $\Sigma$ of a rectangular shape, on the plane defined by two axis of the Cartesian coordinates, let us say $x^{\mu}$ and $x^{\nu}$, with $\mu$ and $\nu$ fixed. The border $\pa\Sigma$ is then the rectangle of infinitesimal sides $\delta x^{\mu}$ and $\delta x^{\nu}$. Evaluating the r.h.s. of \rf{ym_int} in lowest order, and Taylor expanding the integrand around one given corner of the rectangle, we get $\one +  i\,e\(F_{\mu\nu}-\beta\,\widetilde{J}_{\mu\nu}+ie\,\beta^2\sbr{{\widetilde F}_{\mu}}{{\widetilde F}_{\nu}}\)\,\delta x^{\mu}\,\delta x^{\nu}$, with no sum in $\mu$ and $\nu$. Analogously, evaluating the l.h.s. of \rf{ym_int} in lowest order, and Taylor expanding around the  same corner, one gets $\one+\(\pa_{\mu}C_{\nu}-\pa_{\nu}C_{\mu}+\sbr{C_{\mu}}{C_{\nu}}\)\,\delta x^{\mu}\,\delta x^{\nu}$, again with no sum in $\mu$ and $\nu$, and with $C_{\mu}$ given by \rf{cdefym}, and so
\be
\pa_{\mu}C_{\nu}-\pa_{\nu}C_{\mu}+\sbr{C_{\mu}}{C_{\nu}}=ie\,\(F_{\mu\nu} +\beta\,\( D_{\mu}{\widetilde F}_{\nu}-D_{\nu}{\widetilde F}_{\mu}\) +ie\,\beta^2\sbr{{\widetilde F}_{\mu}}{{\widetilde F}_{\nu}}\)
\lab{offshellcurv}
\ee
Therefore, equating both sides of \rf{ym_int}, in lowest order, one gets
\be
D_{\mu}{\widetilde F}_{\nu}-D_{\nu}{\widetilde F}_{\mu}=-\widetilde{J}_{\mu\nu}
\lab{ym_diff_dual}
\ee
where $D_{\mu}*=\pa_{\mu}*+ie\sbr{A_{\mu}}{*}$. 
Taking the Hodge dual one gets the Yang-Mills equations in $(2+1)$ dimensions in the presence of mater currents
\be
D_{\nu}F^{\nu\mu}= J^{\mu}
\lab{ym_diff}
\ee
Note that if one takes the non-abelian Stokes theorem \rf{usual_stokes} with the connection $C_{\mu}$ given by \rf{cdefym}, and so its curvature $G_{\mu\nu}$  given by  \rf{offshellcurv}, one gets the integral equation \rf{ym_int} by using the Yang-Mills equations \rf{ym_diff_dual} . Therefore, \rf{ym_int} is a direct consequence of the non-abelian Stokes theorem \rf{usual_stokes} and the Yang-Mills equations.  In this sense, \rf{ym_int} is an integral formulation of the Yang-Mills theory in (2+1) dimensions in the presence of matter currents.

We have put therefore the Yang-Mills theory in the same footing as the Chern-Simons theory in the presence of matter currents, with its integral equation  being given by \rf{cs_int}. Consequently, most of the results we obtained for Chern-Simons are also valid for the Yang-Mills using the same techniques. For instance, the integral equation \rf{ym_int} transform covariantly under the gauge transformations $A_{\mu}\rightarrow g\,A_{\mu}\,g^{-1}+\frac{i}{e}\,\pa_{\mu}g\,g^{-1}$, and $J_{\mu}\rightarrow g\,J_{\mu}\,g^{-1}$. In addition, it transforms covariantly under re-parameterization of the surface $\Sigma$ with loops, and change of the reference point, as explained in section \ref{sec:cs}. But the most important result following from \rf{ym_int} is that if $\Sigma_c$ is a closed surface with no border, then its l.h.s. is trivial and so
\be
P_{2}\;e^{ie\,\oint_{\Sigma_c}d\tau\;d\sigma\;W^{-1}\left(F_{\mu\nu}-\beta\,\widetilde{J}_{\mu\nu}+ie\,\beta^2\sbr{{\widetilde F}_{\mu}}{{\widetilde F}_{\nu}}\right)\,W\frac{dx^{\mu}}{d\sigma}\frac{dx^{\nu}}{d\tau}}=\one
\lab{closedsigmaym}
\ee
That is the equivalent for Yang-Mills of the equation \rf{charge_2d} for Chern-Simons, and it leads to conservation laws. In particular, for a space-time of the form $M={\cal S}\times \IR$, with ${\cal S}$ being the space sub-manifold, assumed closed with no border, like for instance the two-sphere $S^2$, one gets that $Q_{{\cal S}}=P_{2}\;e^{ie\,\oint_{{\cal S}}d\tau\;d\sigma\;W^{-1}\left(F_{\mu\nu}-\beta\,\widetilde{J}_{\mu\nu}+ie\,\beta^2\sbr{{\widetilde F}_{\mu}}{{\widetilde F}_{\nu}}\right)\,W\frac{dx^{\mu}}{d\sigma}\frac{dx^{\nu}}{d\tau}}=\one$, is constant in time and equal to unity. Again, it implies that the total charge in space  vanishes, or then it may lead to quantization conditions. In the case of an abelian gauge group, for instance $U(1)$, one gets that 
\be
\Phi -\beta\, q=\frac{2\,\pi\,n}{e}\qquad\qquad \mbox{\rm for $n$ integer}
\ee
where $\Phi=\oint_{{\cal S}}d\tau\;d\sigma\; F_{\mu\nu}\,\frac{dx^{\mu}}{d\sigma}\frac{dx^{\nu}}{d\tau}$ is the total magnetic flux (or magnetic charge), and $q=\oint_{{\cal S}}d\tau\;d\sigma\; \widetilde{J}_{\mu\nu}\,\frac{dx^{\mu}}{d\sigma}\frac{dx^{\nu}}{d\tau}$, is the total electric charge in space. 

Again following the reasoning used in the case of the Chern-Simons theory in section \ref{sec:cs}, we get that \rf{closedsigmaym} implies that the quantity 
\be
V\(\Sigma\)=P_{2}\;e^{ie\,\int_{\Sigma}d\tau\;d\sigma\;W^{-1}\left(F_{\mu\nu}-\beta\,\widetilde{J}_{\mu\nu}+ie\,\beta^2\sbr{{\widetilde F}_{\mu}}{{\widetilde F}_{\nu}}\right)\,W\frac{dx^{\mu}}{d\sigma}\frac{dx^{\nu}}{d\tau}}
\lab{sigmaymindep}
\ee
is invariant under smooth deformations  of the surface $\Sigma$ as long as its boundary and reference point $x_R$ are kept fixed. In addition,  it is also invariant under the change of the scanning of $\Sigma$ with loops based at $x_R$. Those facts can be used to construct conserved charges for the Yang-Mills theory. Let us consider space-time to be $\IR^3$, and let us assume that the space components of the field tensor and time component of the currents satisfy the boundary conditions
\be
{\widetilde J}_{12}= J_0 \sim \frac{1}{r^{2+\delta}}\, T\({\hat r}\) \qquad\qquad \qquad
F_{12}\sim \frac{1}{r^{2+\delta^{\prime}}}\, T^{\prime}\({\hat r}\)\qquad\qquad 
{\rm for} \quad r\rightarrow \infty
\ee
with $\delta\, , \, \delta^{\prime}>0$, $r^2=\(x^1\)^2+\(x^2\)^2$, $T\({\hat r}\)$ and $T^{\prime}\({\hat r}\)$ being elements of the Lie algebra of the gauge group $G$, depending on the spatial direction at infinity defined by ${\hat r}=\frac{{\vec r}}{r}$. Consider now a disk ${\cal D}_{\infty}^{(t)}$, of infinite radius on the plane $x^1\,x^2$, at a given time $t$, and let it be scanned with closed loops starting and ending at a reference point $x_R^{(t)}$ on its border. Consider the following operator obtained by integrating \rf{holo_v} on ${\cal D}_{\infty}^{(t)}$, with $G_{\mu\nu}$ given by \rf{gdefym}
\be
V_{x_R^{(t)}}\({\cal D}_{\infty}^{(t)}\)=P_{2}\;e^{ie\,\int_{{\cal D}_{\infty}^{(t)}}d\tau\;d\sigma\;W^{-1}\left(F_{\mu\nu}-\beta\,\widetilde{J}_{\mu\nu}+ie\,\beta^2\sbr{{\widetilde F}_{\mu}}{{\widetilde F}_{\nu}}\right)\,W\frac{dx^{\mu}}{d\sigma}\frac{dx^{\nu}}{d\tau}}=
P_{1}e^{-ie\oint_{S^1_{\infty}}d\sigma\;\left(A_{\mu}+\beta\, {\widetilde F}_{\mu}\right)\,\frac{dx^{\mu}}{d\sigma}}
\lab{eigenconstant_ym}
\ee
where in the last equality we have used the integral equation \rf{ym_int}, and where $S^1_{\infty}$ is the border of ${\cal D}_{\infty}^{(t)}$, i.e. a circle at spatial infinity. 
Then following the arguments used in section \ref{sec:cs}, leading to \rf{eigenconstant_cs}, one concludes that the  eigenvalues of the operator \rf{eigenconstant_ym} are constant in time. Equivalently, one can write those conserved charges as ${\rm Tr}\left[V_{x_R^{(t)}}\({\cal D}_{\infty}^{(t)}\)\right]^N$. Again following those same arguments one concludes that such conserved charges are gauge invariant, and independent of the scanning of ${\cal D}_{\infty}^{(t)}$ with loops, and also on the choice of the reference point $x_R^{(t)}$ on its border.

\subsection{Zero curvature representation for gauge theories}

We now comment on the connection of our integral formulation of Chern-Simons and Yang-Mills theories  in $(2+1)$ dimensions and the approach of \cite{afs,afs-review} using flat connections on  loop spaces. 

Given the Lie algebra-valued 1-form $C=C_{\mu}dx^{\mu}$ and the 2-form $B=\frac{1}{2}B_{\mu\nu}dx^{\mu}\wedge dx^{\nu}$ on $M$, we construct a $1$-form  connection $\mathcal{A}$ in the  loop space $LM=\{ \gamma : S^1\rightarrow M \mid {\rm north}\;{\rm pole} \rightarrow x_R\}$, as \cite{afs,afs-review}
\begin{equation}
\mathcal{A}=\int_{0}^{2\pi}d\sigma\;W^{-1}B_{\mu\nu}W\frac{dx^{\mu}}{d\sigma}\delta x^{\nu}
\end{equation}
where $\delta$ stands for the exterior derivative on the space of all parametrized loops with base point $x_{R}$, with $\delta^{2}=0$ and 
$
\delta x^{\mu}(\sigma)\wedge \delta x^{\nu}(\sigma^{\prime})=-\delta x^{\nu}(\sigma^{\prime})\wedge \delta x^{\mu}(\sigma).
$
The curvature $\mathcal{F}=\delta \mathcal{A}+\mathcal{A}\wedge \mathcal{A}$ is given by
\begin{eqnarray*}
\mathcal{F}\=-\frac{1}{2}\int_{0}^{2\pi}d\sigma\;W(\sigma)^{-1}\left[ D_{\lambda}B_{\mu\nu}+D_{\mu}B_{\nu\lambda}+D_{\nu}B_{\lambda \mu} \right]\lp  x(\sigma) \rp W(\sigma)\frac{dx^{\lambda}}{d\sigma}\delta x^{\mu}(\sigma)\wedge \delta x^{\nu}(\sigma)\\
&+&\frac{1}{2}\int_{0}^{2\pi}d\sigma\;\int_{0}^{2\sigma}d\sigma^{\prime}\;\Bigg( \theta(\sigma -\sigma^{\prime})\left[ B^{W}_{\kappa \mu}\lp x\lp \sigma^{\prime} \rp \rp - G^{W}_{\kappa \mu}\lp x\lp \sigma^{\prime} \rp \rp ,B^{W}_{\lambda \nu}\lp x\lp \sigma \rp \rp \right]\\
&-&\theta(\sigma^{\prime} -\sigma)\left[ B^{W}_{\lambda \nu}\lp x\lp \sigma \rp \rp - G^{W}_{\lambda \nu}\lp x\lp \sigma \rp \rp ,B^{W}_{\kappa \mu}\lp x\lp \sigma^{\prime} \rp \rp \right]  \Bigg)\frac{dx^{\kappa}}{d\sigma^{\prime}}\frac{dx^{\lambda}}{d\sigma}\delta x^{\mu}(\sigma^{\prime})\wedge \delta x^{\nu}(\sigma)
\end{eqnarray*}
where $G_{\mu\nu}$ is the curvature of $C_{\mu}$, i.e. $G_{\mu\nu}=\pa_{\mu}C_{\nu}-\pa_{\nu}C_{\mu}+[C_{\mu},C_{\nu}]$, and $W$ is constructed out of $C_{\mu}$ by integration of \rf{holo_w}.
 
In the case of Chern-Simons theory we consider 
$$
C_{\mu}=ieA_{\mu}\quad 
\text{and}\quad 
B_{\mu\nu}=\frac{1}{\kappa}{\widetilde J}_{\mu\nu}.$$
Lemma $2.1$ of \cite{afs-review} claims that if
$$
B_{\mu\nu}-G_{\mu\nu}=0
$$ 
then $\mathcal{F}=0$. Then, for our choice of $C_{\mu}$ and $B_{\mu\nu}$ given above we see that 
$$
\mathcal{F}=0 \Leftrightarrow \text{Chern-Simons equation is satisfied}.
$$
For Yang-Mills theory we take 
$$C_{\mu}=ie \lp A_{\mu}+\beta {\widetilde F}_{\mu}  \rp \quad \text{and} \quad 
B_{\mu\nu}=i\,e\(F_{\mu\nu}-\beta\,\widetilde{J}_{\mu\nu}+ie\,\beta^2\sbr{{\widetilde F}_{\mu}}{{\widetilde F}_{\nu}}\).$$
The condition $B_{\mu\nu}-G_{\mu\nu}=0$ leads to the Yang-Mills equation 
$ D_{\nu}F^{\nu\mu}=J^{\mu}$, and therefore gives the zero curvature representation of this theory in loop space.

Therefore, for the cases of Chern-Simons and Yang-Mills theories in $(2+1)$ dimensions, the   integral formulation approach  and zero curvature on loop space lead to the same results, and also to the same conserved charges.

\section{The case of volumes: theories in $3+1$ dimensions}
\label{sec:3+1}
\setcounter{equation}{0}

For $3+1$ dimensional theories we consider the generalization of the non-abelian Stokes theorem for a 2-form connection proved in appendix B. Given a three dimensional volume $\Omega$ and its two dimensional border $\partial \Omega$, a closed surface, the theorem relates the surface-ordered integral of the connection $W^{-1}B_{\mu\nu}W$ along $\partial \Omega$ with the volume-ordered integral of 
\begin{eqnarray*}
\mathcal{K}\=\int_{0}^{2\pi}d\tau V\Bigg\{\int_{0}^{2\pi}d\sigma\;W^{-1}\lp D_{\rho}B_{\mu\nu}+D_{\mu}B_{\nu\rho}+D_{\nu}B_{\rho\mu} \rp W\frac{dx^{\mu}}{d\sigma}\frac{dx^{\nu}}{d\tau}\frac{dx^{\rho}}{d\zeta}+\\
&-&\int_{0}^{2\pi}d\sigma \int_{0}^{\sigma}d\sigma^{\prime}\;\lc B^{W}_{\kappa \lambda}(\sigma^{\prime})-F^{\mu\nu}_{\kappa \lambda}(\sigma^{\prime}),B^{W}_{\mu\nu}(\sigma) \rc \frac{dx^{\kappa}}{d\sigma^{\prime}}(\sigma^{\prime})\frac{dx^{\mu}}{d\sigma}(\sigma)\times \\
&\times &\lp  \frac{dx^{\lambda}}{d\tau}(\sigma^{\prime})\frac{dx^{\nu}}{d\zeta}(\sigma)-
\frac{dx^{\lambda}}{d\zeta}(\sigma^{\prime})\frac{dx^{\nu}}{d\tau}(\sigma)  
\rp\Bigg\}V^{-1}
\end{eqnarray*}
in $\Omega$:
\begin{equation}
\lab{stokes_3d}
V_R\, P_2 e^{\int_{\partial\Omega}d\tau d\sigma W^{-1}B_{\mu\nu}W \frac{dx^{\mu}}{d\sigma}\,\frac{d\,x^{\nu}}{d\,\tau}}= P_3e^{\int_{\Omega} d\zeta \,{\cal K}}\,V_{R}.
\end{equation}
In the next few lines we summarize the construction of the above equation. A reference point $x_{R}$ is defined on the border of $\Omega$, and around it we construct an infinitesimal volume, whose border is the infinitesimal closed surface $\Sigma_{R}$. We then scan $\Omega$ with closed surfaces based at the reference point $x_R$.  In doing so, we define the parameter $\zeta$, such that $\zeta = 0 $ stands for the infinitesimal surface $\Sigma_{R}$ and $\zeta = 2\pi$, for the boundary $\partial \Omega$. During this variation the quantities $V$ are calculated for each surface, in each step, through equation
\begin{equation}
\label{holo_v_tau}
\frac{dV}{d\tau}-V\mathcal{A}=0
\end{equation}
with $\mathcal{A}=\int_{0}^{2\pi}d\sigma W^{-1}B_{\mu\nu}W \frac{dx^{\mu}}{d\sigma}\frac{dx^{\nu}}{d\tau}d\sigma $. The surface $\Sigma$ is scanned with loops parametrized by $\sigma \in [0,2\pi]$, starting and ending at the reference point, and the parameter $\tau$ labels these loops. The Wilson lines $W$ are calculated on the loops using equation $\eqref{holo_w}$. The l.h.s of \rf{stokes_3d} is therefore obtained after integrating \eqref{holo_v_tau} over the surface $\partial \Omega$, with the ordering given by the way we scan it with loops. The quantity $V_{R}$ is the integration constant (the value of $V$ for the infinitesimal surface $\Sigma_{R}$).

Once the surface $\Sigma$ is closed, $V$ can also be obtained from the equation (as we show in the appendix)
\begin{equation}
\label{holo_v_zeta}
\frac{dV}{d\zeta}-\mathcal{K}V=0
\end{equation}
and this result is expressed in the r.h.s of the theorem above.

We now proceed to show how to formulate an integral version of Yang-Mills theory through this non-abelian Stokes theorem, and also, how it leads to conserved charges.

\subsection{Integral formulation of Yang-Mills theory with source}

We consider the Yang-Mills theory in $(3+1)$ dimensions for a gauge group $G$ and in the presence of matter currents $J_{\mu}$. The classical equations of motion are given by
\be
D_{\nu}{\widetilde F}^{\nu\mu}=0 \qquad\qquad \qquad D_{\nu}F^{\nu\mu}= J^{\mu}
\lab{ym_diff_3p1}
\ee
where $F_{\mu\nu}=\partial_{\mu} A_{\nu}-\partial_{\nu} A_{\mu} + i\,e\,\sbr{A_{\mu}}{A_{\nu}}$, and  ${\widetilde F}_{\mu\nu}$ is the Hodge dual of the field tensor, \emph{i.e.}, $F_{\mu\nu}\equiv \frac{1}{2}\,\varepsilon_{\mu\nu\rho\lambda}\, {\widetilde F}^{\rho\lambda}$.

One can obtain an integral equation for the Yang-Mills theory using the Stokes theorem \rf{stokes_3d} as follows (see \cite{first} for more details). Take $B_{\mu\nu}=ie\left[ \alpha\, F_{\mu\nu}+\beta\, {\widetilde F}_{\mu\nu}\right]$, with $\alpha$ and $\beta$ being arbitrary constants, and using Yang-Mills differential equations \rf{ym_diff_3p1}  to replace $D_{\rho}B_{\mu\nu}+D_{\mu}B_{\nu\rho}+D_{\nu}B_{\rho\mu}$ by $(-ie\beta {\widetilde J}_{\mu\nu\lambda})$. With that, the quantity ${\cal K}$ above is now given by ${\cal K}=\int_0^{2\,\pi}d\tau\, V\, {\cal J}\,V^{-1}$, 
with 
\begin{eqnarray*}
\mathcal{J}&\equiv & \int_{0}^{2\pi}d\sigma\;\Bigg\{  ie\beta W^{-1}\widetilde{J}_{\mu\nu\rho}W\;\frac{dx^{\mu}}{d\sigma}\frac{dx^{\nu}}{d\tau}\frac{dx^{\lambda}}{d\zeta}+\\ \nonumber
&+& e^{2}\int_{0}^{\sigma} d\sigma^{\prime}\left[ \lp (\alpha-1)F^{W}_{\kappa\rho}+\beta \widetilde{F}^{W}_{\kappa\rho} \rp(\sigma^{\prime}), \lp \alpha F^{W}_{\mu\nu}+\beta \widetilde{F}^{W}_{\mu\nu} \rp (\sigma)\right]\times \\ \nonumber
&\times & \frac{dx^{\kappa}}{d\sigma^{\prime}} \frac{dx^{\mu}}{d\sigma}\lp \frac{dx^{\rho}(\sigma^{\prime})}{d\tau} \frac{dx^{\nu}(\sigma)}{d\zeta}- \frac{dx^{\rho}(\sigma^{\prime})}{d\zeta} \frac{dx^{\nu}(\sigma)}{d\tau}\rp \Bigg\}
\end{eqnarray*}
and $V$ is constructed by integrating  \eqref{holo_v_tau} with $\mathcal{A}\equiv ie\,
 \int_{0}^{2\pi}d\sigma W^{-1}\left[ \alpha F_{\mu\nu}+\beta{\widetilde F}_{\mu\nu}\right] W \frac{dx^{\mu}}{d\sigma}\frac{dx^{\nu}}{d\tau}$, and where we have used the notation $X^W\equiv W^{-1}\,X\,W$. Then we have \cite{first}:
 \begin{equation}
 \lab{ym_int_4}
 P_2 e^{ie\int_{\partial\Omega}d\tau d\sigma \left[ \alpha F_{\mu\nu}^W+\beta {\widetilde F}_{\mu\nu}^W\right] \frac{dx^{\mu}}{d\sigma}\frac{dx^{\nu}}{d\tau}}=  P_3e^{\int_{\Omega} d\zeta d\tau  V{\cal J}V^{-1}}.
\end{equation}
which is a direct consequence of the generalized non-abelian Stokes theorem \rf{stokes_3d} and the Yang-Mills equations \rf{ym_diff_3p1}. On the other hand, the integral equation \rf{ym_int_4} implies the differential equations \rf{ym_diff_3p1}, as we now explain.

\begin{figure}
\begin{center}
\includegraphics[scale=0.3]{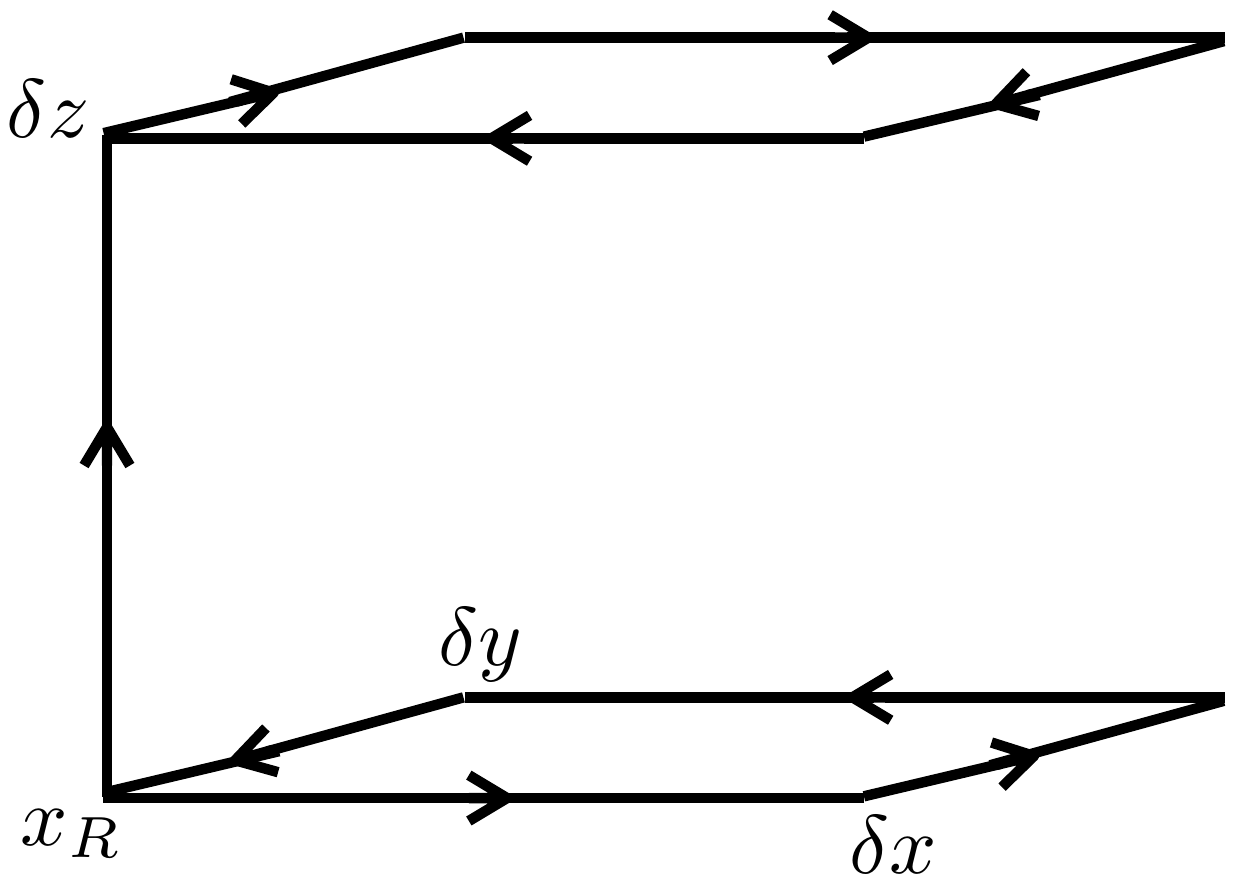}
\caption{The scanning of an infinitesimal cube.}
\label{cube_scan}
\end{center}
\end{figure}

Equation \rf{ym_int_4} is defined for an arbitrary volume $\Omega$, and in particular, for an infinitesimal one.  Take $\Omega$ as an infinitesimal cube of sides $\delta x^ {\mu}$, $\delta y^{\nu}$ and $\delta z^{\lambda}$, with the indices fixed (see Figure \ref{cube_scan}). We choose the reference point $x_{R}$ to be at one of the vertices, and when opposite surfaces are scanned with loops based on $x_{R}$ one has to pay special attention to the fact that the signs of the velocities in $\tau$-direction changes. 

Now, considering only first order contributions to equation \rf{ym_int_4}, the integrand in its l.h.s can be evaluated at any point on the cube's face, since the differences will be of higher order, and therefore, evaluating it on the face $\delta z^{\lambda} = 0$  one gets\footnote{The minus sign is due to the choice of the direction of scanning.} $-ie\left[\alpha F_{\mu\nu}+\beta{\widetilde F}_{\mu\nu}\right]_{x_R}\delta x^{\mu}\delta y^{\nu}$. For the face at $x_{R}+\delta z$ the contribution comes from $ie\lp W^{-1}\left[\alpha F_{\mu\nu}+\beta{\widetilde F}_{\mu\nu}\right]W\rp_{\lp x_R+\delta z^{\lambda}\rp}\delta x^{\mu}\delta y^{\nu}$. For this term we have to expand both the Wilson line and the field strength so that their values at $x_{R}+\delta z$ are approximated to their values in $x_{R}$. The equation for the Wilson line gives for an infinitesimal variation $\sigma \rightarrow \sigma + \delta \sigma$ along the $z$ direction, $W_{\lp x_R+\delta z^{\lambda}\rp }\sim1-ieA_{\lambda}\lp x_R\rp \delta x^{\lambda}$ and Taylor expanding the field strength in this direction, $F_{\mu\nu}(x_{R}+\delta z)=F_{\mu\nu}(x_{R})+\pa_{\lambda}F_{\mu\nu}(x_{R})\delta z^{\lambda}$, we end up with $ieD_{\lambda}\left[\alpha F_{\mu\nu}+\beta{\widetilde F}_{\mu\nu}\right]_{x_R}\delta x^{\mu}\delta y^{\nu}\delta z^{\lambda}$. Doing the same for the other two pairs of faces one gets
$$
P_2 e^{ie\int_{\partial\Omega}d\tau d\sigma \left[ \alpha F_{\mu\nu}^W+\beta {\widetilde F}_{\mu\nu}^W\right] \frac{dx^{\mu}}{d\sigma}\frac{dx^{\nu}}{d\tau}}\approx \one +ie(D_{\lambda}[\alpha F_{\mu\nu}+\beta{\widetilde F}_{\mu\nu}]+\mbox{cyclic perm.})_{x_R}\delta x^{\mu}\delta y^{\nu}\delta z^{\lambda}.
$$
For the r.h.s of \rf{ym_int_4}, considering only the first order contributions, we notice that the commutator term is of higher order with respect to the first term involving only the current, for it has one more integration along the loop. Then, up to lowest order 
$$
 P_3e^{\int_{\Omega} d\zeta d\tau  V{\cal J}V^{-1}} \approx \one+ie\beta{\widetilde J}_{\mu\nu\lambda}\delta x^{\mu}\delta y^{\nu}\delta z^{\lambda}
$$
and clearly equating the previous result with this one, we get the set of Yang-Mills equations \rf{ym_diff_3p1}, as the coefficients of the parameters $\alpha$ and $\beta$. 

Comparing  the integral equation \rf{ym_int_4}  with \rf{stokes_3d} one notices that the integration constants $V_{R}$ are missing. One has to keep in mind that while \rf{stokes_3d} is a mathematical relation, \rf{ym_int_4} is a physical equation, and therefore gauge covariance is important, and in order to guarantee that  those integration constants must lie in the center of the gauge group, as we now explain  in detail. Consider the gauge transformation of the Yang-Mills field $A_{\mu}\rightarrow g A_{\mu}g^{-1}+\frac{i}{e}\pa_{\mu}gg^{-1}$, which implies the following transformations of the field strength $F_{\mu\nu}\rightarrow gF_{\mu\nu}g^{-1}$, and the matter current transforms in the same way, $J_{\mu}\rightarrow gJ_{\mu}g^{-1}$. From \eqref{holo_w}, the Wilson line changes as $W \rightarrow g_{f}Wg^{-1}_{i}$ where $g_{i}$ and $g_{f}$ stand for the value the gauge group element takes at the points $x_{i}$ and $x_{f}$ respectively; for the reference point we denote as $g_{R}$ the value of the gauge group element there. Then it is direct to see (by replacing $W$, $F_{\mu\nu}$ and $J_{\mu}$ by the respective gauge transformed quantities given before) that all the important quantities related to the integral Yang-Mills equation \rf{ym_int_4} (such as $V$, $\mathcal{J}$, $\mathcal{K}$ and $\mathcal{A}$) transform as $V\rightarrow g_{R}Vg_{R}^{-1}$, $\mathcal{J}\rightarrow g_{R}\mathcal{J}g_{R}^{-1}$, $\mathcal{K}\rightarrow g_{R}\mathcal{K}g_{R}^{-1}$ and $\mathcal{A}\rightarrow g_{R}\mathcal{A}g_{R}^{-1}$. 

The l.h.s of \rf{ym_int_4} comes from the l.h.s of the non-abelian Stokes theorem previously presented, which, in turn, is the result of the integration of equation \eqref{holo_v_tau}. We notice that if $V$ is a solution of this equation, then $V^{\prime}=kV$ is also a solution, where $k$ is a constant element of the gauge group. Under a gauge transformation one gets $V^{\prime}\rightarrow g_{R}V^{\prime} g_{R}^{-1}=g_{R}k  V g_{R}^{-1}$. On the other hand, since $k$ is an arbitrary constant, and so insensitive to transformations of the gauge field, one  could have written it as $V^{\prime}\rightarrow kg_{R}V g_{R}^{-1}$, and the only way to guarantee the compatibility is to have $g_{R}k=kg_{R}$, \emph{i.e.}, to have $k$ in the center $Z(G)$ of the group. The same argument can be applied to the r.h.s of \rf{ym_int_4}, which comes from the r.h.s of the non-abelian Stokes theorem, which, in turn, is the solution of equation \eqref{holo_v_zeta}. In that case, if $V$ is a solution, so is $V^{\prime}=Vh$, with $h$ a constant element of $G$. Finally we conclude that  \rf{ym_int_4} is gauge covariant only if the integration constants are in the center of the gauge group. Then they can be cancelled trivially, since they commute with the surface and volume integrals, and that is the reason why they do not appear in the integral Yang-Mills equation \rf{ym_int_4}.

An important issue concerns the fact that equation \rf{ym_int_4} is formulated in a way that it depends on the particular choice of the reference point $x_{R}$ and of the scanning of the volume with surfaces. Although it is at first sight unwanted, the Stokes theorem \rf{stokes_3d} guarantees that if one changes any of these things, each side of \rf{ym_int_4} will change in a way to remain equal to each other; in this sense, this equation transforms ``covariantly'' under reparametrization. This can be better understood once we realize that this equation is formulated in the loop space $L\Omega =\{ \gamma: \mathcal{S}^{2}\rightarrow \Omega \vert \text{north pole} \rightarrow x_{R} \in \partial \Omega\}$, formed by maps from $\mathcal{S}^{2}$ into $\Omega$, such that the north pole of $\mathcal{S}^{2}$ is mapped into the reference point $x_{R}$. The images of this map are closed surfaces in $\Omega$, starting and ending at $x_{R}$. The volume $\Omega$ is scanned by a family of these closed surfaces, which are points in $L\Omega$, thus, $\Omega$ is a path in the loop space $L\Omega$. Then, a change in the scanning of $\Omega$ corresponds to a change on the parametrization of the path in $L\Omega$, which does not change the path, nor the physical results from it.

Equation \rf{ym_int_4} does not only describes the Yang-Mills theory in loop space, but also leads to a conservation laws as we now discuss. For a closed path in loop space, corresponding to a closed volume $\Omega_{\text{c}}$ in space-time, with no boundary, the l.h.s of equation \rf{ym_int_4} becomes trivial and we get
 \begin{equation}
 \label{charge_4d}
 P_3e^{\int_{\Omega_{\text{c}}} d\zeta d\tau  V{\cal J}V^{-1}}=\one.
\end{equation}
Consider now a given point in that path, in space-time, the surface $\Sigma$, which divides it into two parts: $\Omega_{\text{c}}=\Omega_{1}+\Omega_{2}$. Then, we can split equation \eqref{charge_4d} as
\begin{equation}
\label{charge_4d_indep}
P_3e^{\int_{\Omega_{2}} d\zeta d\tau  V{\cal J}V^{-1}}\cdot P_3e^{\int_{\Omega_{1}} d\zeta d\tau  V{\cal J}V^{-1}}=\one.
\end{equation}
Each term of this equation is obtained from integration of \eqref{holo_v_zeta}. It can be written as $V(\Omega_{2})\cdot V(\Omega_{1})=\one$. Now, reverting the order of integration for the second path $\Omega_{2}$, one gets equivalently $V^{-1}(\Omega_{2}^{-1})\cdot V(\Omega_{1})=\one$, which implies that $V(\Omega_{1})=V(\Omega_{2}^{-1})$. In other words, the  integration of \eqref{holo_v_zeta}, for the volumes $\Omega_{1}$ and $\Omega_{2}^{-1}$, with the same boundary $\Sigma$ and reference point $x_R$, leads to the same operator. In fact, since equation \eqref{charge_4d_indep} holds for any closed volume $\Omega_{\text{c}}$ and for any partition of it into two other volumes, what we just saw is that not only the integration of  \eqref{holo_v_zeta} along two volumes with same boundary gives the same operator but also that $P_3e^{\int_{\Omega} d\zeta d\tau  V{\cal J}V^{-1}}$ is independent of the volume, for any volume $\Omega$. This means that as long as the border is kept fixed, one can change the volume and there will be no consequences to $V(\Omega)$. The fact that this operator depends only on the border $\partial \Omega$ and on the reference point $x_{R}$ on it, leads to conservation laws as we now explain.

Consider first the case in which space-time $M$ has the topology $\mathcal{S}\times \IR$, with $\mathcal{S}$ being a closed unbounded spatial submanifold and $\IR$ the time. Then, equation \eqref{charge_4d} can be evaluated for $\Omega_{\text{c}}=\mathcal{S}$, giving that  the quantity
\begin{equation}
\label{charge_closed_4}
Q_{{\cal S}}\equiv P_3e^{\oint_{{\cal S}} d\zeta d\tau  V{\cal J}V^{-1}}=\one.
\end{equation}
is conserved in time and is equal to the unit. This has two main implications: the first is that the net charge on the whole space vanishes, and the other is that it might lead to some quantization condition. In particular, for the case of the Maxwell theory, where the (abelian) gauge group is $G=U(1)$, this equation for $Q_{{\cal S}}$ is satisfied if  
$q\equiv \int_{{\cal S}}d\zeta d\tau d\sigma{\widetilde J}_{\mu\nu\lambda}
\frac{dx^{\mu}}{d\sigma}\frac{dx^{\nu}}{d\tau}
\frac{dx^{\lambda}}{d\zeta}$, is such that
$$
q= \frac{2\pi n}{e\beta}
$$ 
with $n$  integer. If for some reason $\beta$ can be fixed  at the quantum level, then \eqref{charge_closed_4} express the quantization of electric charge \cite{alvarez-olive}. 

\begin{figure}[t]
  \begin{center}
    \includegraphics[width=0.5\textwidth]{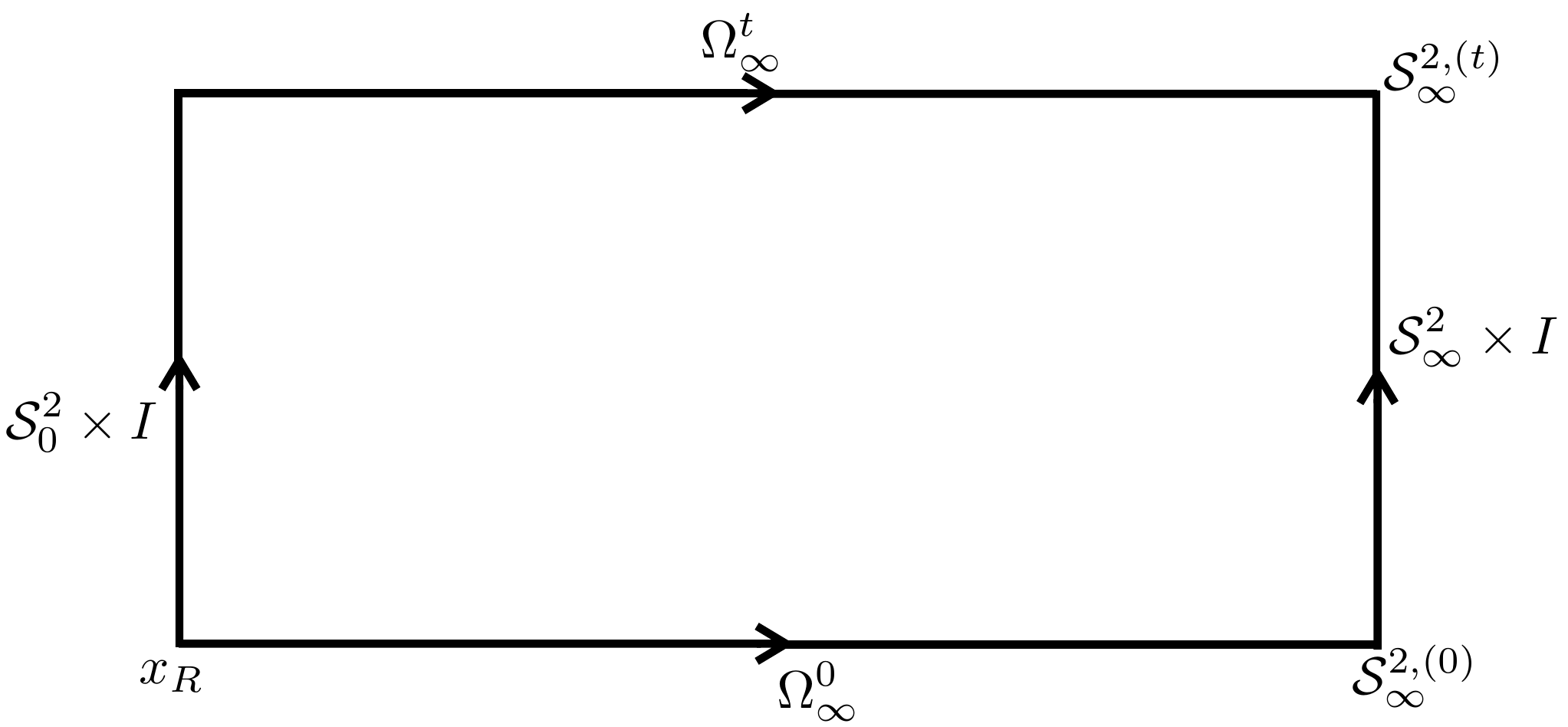}
\caption{The schematic representation in loop space $L\Omega$} of the ``path'' used in the construction of the conserved charge.  
\label{path_charge_3d}
  \end{center}
\end{figure}

Next we consider the space-time $M$ to have the topology $\IR^{3}\times \IR$. Lets take two points in the loop space, which correspond in space-time to the following two surfaces (see figure \ref{path_charge_3d}): the infinitesimal 2-sphere $\mathcal{S}_{0}^{2}$ around $x_{R}$, at time $x^{0}=0$ and the 2-sphere which is  the border of the entire space at time $x^{0}=t$; and which we denote as $\mathcal{S}_{\infty}^{2, (t)}$. Now, these two points in $L\Omega$ are joined by two different paths (volumes in space-time) and each of these paths are composed by two parts as we now explain. The first part of the first path is the infinitesimal hyper-cylinder $\mathcal{S}_{0}^{2}\times I$, with $I$ being the interval in $\IR$ from $0$ to $t$ and $\mathcal{S}_{0}^{2}$ the infinitesimal sphere around $x_{R}$. The second part of this path is the volume inside the sphere $\mathcal{S}_{\infty}^{2, (t)}$, which we make by ``blowing up'' the infinitesimal sphere $\mathcal{S}_{0}^{2}$ when we reach the point $x^{0}=t$; we denote this volume by $\Omega_{\infty}^{(t)}$. This first path is then the composition $\Omega_{1}=(\mathcal{S}_{0}^{2}\times I )\cup \Omega_{\infty}^{(t)}$. The second path has a first part, which in space-time corresponds to the whole spatial volume $\Omega_{\infty}^{(0)}$, inside $\mathcal{S}_{\infty}^{2, (0)}$ at $x^{0}=0$, and its second part is the hyper-cylinder $\mathcal{S}_{\infty}^{2}\times I$. This path is the composition $\Omega_{2}=\Omega_{\infty}^{(0)} \times (\mathcal{S}_{\infty}^{2}\times I)$. Basically we are dealing with different paths (volumes) which boundaries - $\mathcal{S}_{0}^{2}$ and  $\mathcal{S}_{\infty}^{2,(t)}$ - are common, and thus, following our previous result, the operator $V(\Omega)= P_3e^{\int_{{\Omega}} d\zeta d\tau  V{\cal J}V^{-1}}$, is independent of the volume, and  should depend only on the boundaries of $\Omega$. Therefore, it follows that $V(\Omega_{2})=V(\Omega_{1})$, and this can be written as 
\begin{equation}
\label{path_indep}
V(\mathcal{S}_{\infty}^{(2)}\times I)V(\Omega_{\infty}^{(0)})=V(\Omega_{\infty}^{(t)})V(\mathcal{S}_{0}^{2}\times I).
\end{equation} 
In order to obtain $V(\Omega)$ for each step it is necessary to evaluate $\mathcal{K}=\int_{0}^{2\pi}d\tau \;V\mathcal{J}V^{-1}$ on the surfaces scanning each volume $\Omega$. We shall scan a hyper-cylinder $S^2\times I$ with surfaces, based at $x_R$, of the form given in figure (5.b), with $t^{\prime}$ denoting a time in the interval $I$. Each one of such surfaces are scanned with loops, labelled by $\tau$, in the following way. For $0\leq \tau\leq \frac{2\pi}{3}$, we scan the infinitesimal cylinder as shown in figure (5.a), then for $\frac{2\pi}{3}\leq \tau\leq \frac{4\pi}{3}$ we scan the sphere $S^2$ as shown in figure (5.b), and finally for $\frac{4\pi}{3}\leq \tau\leq 2\pi$ we go back to  $x_R$ with loops as shown in figure (5.c). Then, we can split $\mathcal{K}$ in three parts: $\mathcal{K}_{a}+\mathcal{K}_{b}+\mathcal{K}_{c}$, each of them corresponding to one of the three surfaces, defined by the $\tau$ intervals. 

From the physical point of view it is very reasonable to take the current and the field strength to satisfy the boundary conditions 
$$
J_{\mu}\sim \frac{1}{R^{2+\delta}} \qquad \qquad F_{\mu\nu}\sim \frac{1}{R^{\frac{3}{2} + \delta^{\prime}}}
$$
for $R\rightarrow \infty$, with $\delta$ and $\delta^{\prime}$ bigger than zero. With that, integration of $\mathcal{J}$ over $\mathcal{S}_{\infty}^{2}$ vanishes and we get $\mathcal{K}_{b}=0$ for the path $\Omega_{2}$. We notice that for the path $\Omega_{1}$, the infinitesimal sphere $\mathcal{S}_{0}^{2}$ does not contribute to $\mathcal{K}_{b}$, and we have also $\mathcal{K}_{b}=0$ in this case. Then we conclude that $\mathcal{K}$ calculated in both spheres $\mathcal{S}_{\infty}^{2}$ and $\mathcal{S}_{0}^{2}$ gives the same result and therefore
$$
V(\mathcal{S}_{\infty}^{(2)}\times I)=V(\mathcal{S}_{0}^{2}\times I).
$$
It is now possible to contract the cylinders into a line, so that the loops scanning them in the intervals $\tau \in [0,\frac{2\pi}{3}]$ and $\tau \in [\frac{4\pi}{3},2\pi]$ become exactly the same. However, since one set of loops is ``going up'', and the other is ``going down'' (the sign of the velocity $\frac{dx^{\mu}}{d\tau}$ changes), the contributions $\mathcal{K}_{a}$ and $\mathcal{K}_{c}$ exactly cancel. Note that $V$ inside $\mathcal{K}$, obtained from \eqref{holo_v_tau} does not change sign since it does not see the direction the loops are going to but only takes into account the profile of the loop for each $\tau$.

\begin{figure}[ht]
\centering
\subfigure[]{
   \includegraphics[width=0.059\textwidth]{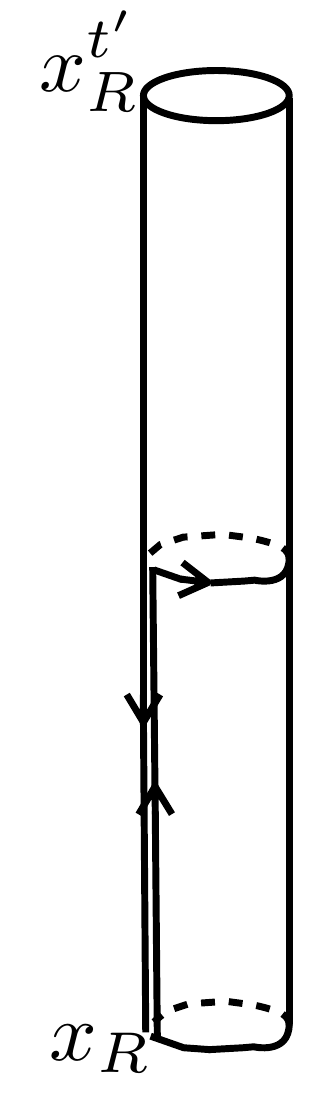}
   \label{closed_surface_a}
} \hspace{1cm}
 \subfigure[]{
   \includegraphics[width=0.22\textwidth]{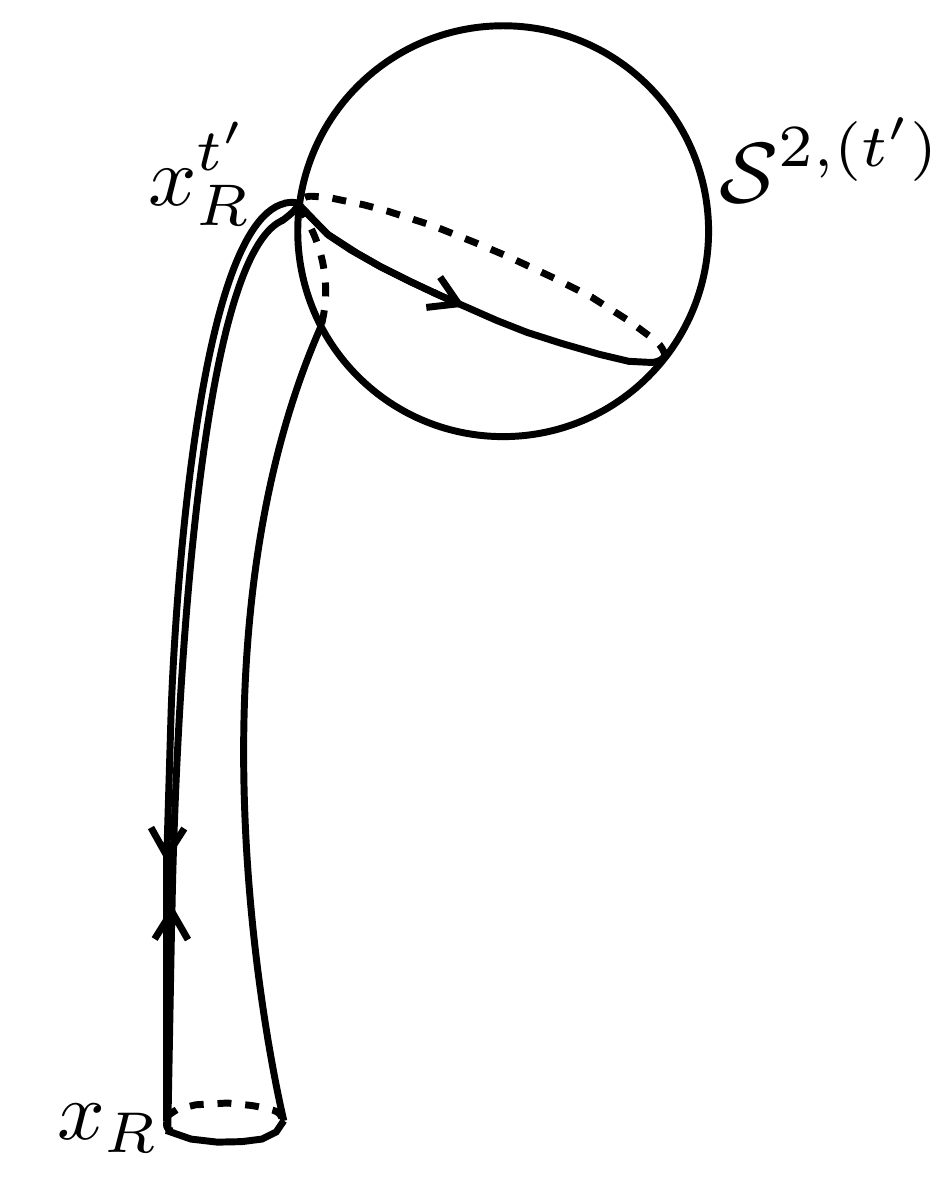}
   \label{closed_surface_b}
 } \hspace{1cm}
 \subfigure[]{
   \includegraphics[width=0.061\textwidth]{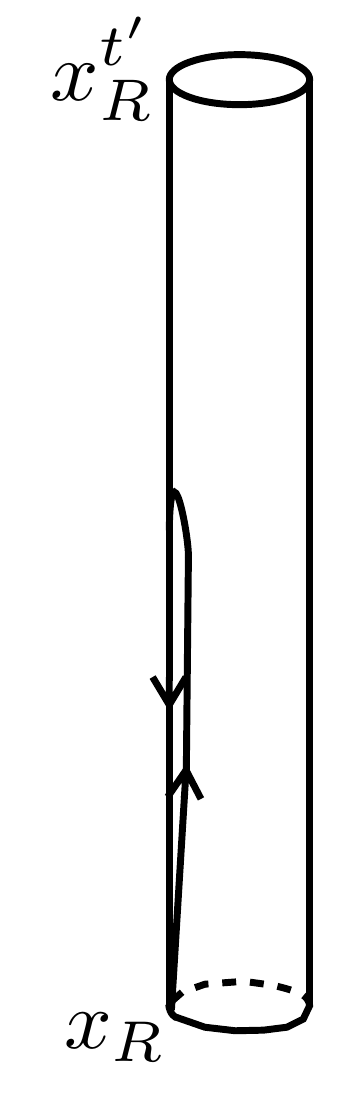}
   \label{closed_surface_c}
 }
\caption{Surfaces of type (b) scan a hyper-cylinder $S^2\times I$.}
\end{figure}

Finally what remains is exactly $V$ calculated on $\Omega_{\infty}^{(t)}$, whose border is the sphere $\mathcal{S}_{\infty}^{2,(t)}$, where we need to evaluate $\mathcal{K}$. In order to scan the sphere, we establish the point $x_{R}^{(t)}$, on its boundary. This is the point $x_{R}$ at time $x^{0}=t$, so that the path starts at $x_{R}$, goes up to $x_{R}^{(t)}$, and from this point we scan the sphere with loops. Then, we go back to $x_{R}$. So, we construct the Wilson line composed of the two parts: $W=W(x,x_{R}^{(t)})W(x_{R}^{(t)},x_{R})$ the one corresponding to the ``leg'' that goes up from $x_{R}$ to $x_{R}^{(t)}$, denoted by $W(x_{R}^{(t)},x_{R})$, and the one that corresponds to the path from $x_{R}^{(t)}$ to the point $x(\sigma)$, on the sphere. Using this decomposition we can re-write every needed quantity in terms of the new reference point $x_{R}^{(t)}$. In particular, $\mathcal{A}$ in equation \eqref{holo_v_tau} is decomposed as $\mathcal{A}_{x_{R}}=W(x_{R}^{(t)},x_{R})^{-1}\mathcal{A}_{x_{R}^{(t)}}W(x_{R}^{(t)},x_{R})$. This leads to $V_{x_{R}}=W^{-1}(x_{R}^{(t)},x_{R}) V_{x_{R}^{(t)}}W(x_{R}^{t},x_{R})$;  In the same way the quantities $\mathcal{K}_{b}$ and $\mathcal{J}$ transform as $\mathcal{K}_{b,x_{R}}=W^{-1}(x_{R}^{(t)},x_{R}) \mathcal{K}_{b,x_{R}^{(t)}}W(x_{R}^{t},x_{R})$ and $\mathcal{J}_{x_{R}}=W^{-1}(x_{R}^{(t)},x_{R}) \mathcal{J}_{x_{R}^{(t)}}W(x_{R}^{t},x_{R})$, so that $V(\Omega)$ obtained from \eqref{holo_v_zeta} becomes
$$
V_{x_{R}}(\Omega_{\infty}^{(t)})=W^{-1}(x_{R}^{(t)},x_{R})V_{x_{R}^{t}}(\Omega_{\infty}^{(t)}) W(x_{R}^{(t)},x_{R}).
$$
Plugging this into the equation \eqref{path_indep} we finally get
\begin{equation}
V_{x_{R}^{t}}(\Omega_{\infty}^{(t)})=U(t)\cdot V(\Omega_{\infty}^{(0)})\cdot U^{-1}(t)
\end{equation}
with $U(t)=W(x_{R}^{(t)},x_{R})\cdot V(\mathcal{S}^{2}_{0}\times I)$.

Therefore, the operator $V_{x_{R}^{(t)}}(\Omega_{\infty}^{(t)})$ has an iso-spectral evolution in time, and thus, its eigenvalues, or equivalently $\text{Tr}\lp V_{x_{R}}(\Omega_{\infty}^{(t^{\prime})}) \rp^{N}$, are conserved in time. Using Yang-Mills integral equation one can write $V_{x_{R}^{t}}(\Omega_{\infty}^{(t)})$ as a volume or a surface ordered integral:
\begin{equation}
V_{x^{(t)}_{R}}(\Omega_{\infty}^{(t)})=P_{2}\;e^{ie \int_{\mathcal{S}_{\infty}^{2,(t)}}d\tau\;d\sigma\;\lp \alpha F^{W}_{\mu\nu}+\beta {\widetilde F}_{\mu\nu}^{W} \rp \frac{dx^{\mu}}{d\sigma}\frac{dx^{\nu}}{d\tau}}=P_{3}\;e^{\int_{\Omega_{\infty}^{(t)}}d\zeta\;d\tau\;V\mathcal{J}V^{-1}}.
\end{equation}

The fact that this operator is independent of the parameterization of the volume guarantees that the conserved charges constructed from it are also independent of this parametrization. One can, however, argue about the dependence of our results with respect to the choice of the reference point. As we saw above, by changing $x_{R}^{(t)}$ to ${\widetilde x_{R}^{t}}$, $V_{x_{R}^{(t)}}$ changes under conjugation with respect to $W({\widetilde x_{R}^{t}},x_{R}^{t})$, and so its eigenvalues (conserved charges) are invariant under change of reference point. In addition, the reference point is placed at the border of $\Omega_{x_{R}^{t}}$ (the spatial infinity), and the field strength is supposed to vanish there, which implies that the gauge field $A_{\mu}$ is asymptotically flat, and therefore the Wilson line depends only on the points and not on the path joining them. The conserved charges are also gauge invariant, since under a gauge transformation we have seen that $V_{x_R^{(t)}}(\Omega_{\infty}^{(t)})\rightarrow g_RV_{x_R^{(t)}}(\Omega_{\infty}^{(t)})g_R^{-1}$, with $g_R$ being  the group element, performing the gauge transformation, at $x_R^{(t)}$. Note in addition that if $V_{x_R^{(t)}}(\Omega_{\infty}^{(t)})$ has an iso-spectral evolution so does $g_cV_{x_R^{(t)}}(\Omega_{\infty}^{(t)})$, with $g_c$ and element of the centre  $Z(G)$ of the gauge group $G$. That fact has to do with the freedom we have to choose the integration constants  to lie in $Z(G)$, without spoiling the gauge covariance of \rf{ym_int_4}.

Note that there are several integral and loop space formulations of Yang-Mills theories \cite{loopgauge}. Our approach shares some of the ideas of those formulations in the sense of using ordered integrals of the gauge potential and of the field strength tensor. However, it differs in an essential way because  it is based on the new eq. \rf{ym_int_4}. In addition, it leads in a quite novel and direct way to gauge invariant conserved charges.

\subsection{The conserved charges for Dyons}

Consider a theory where the gauge group $G$ is spontaneously broken into a Higgs field in the adjoint representation. Consider a BPS dyon solution:
$$
E_{i}=-F^{0i}=\sin{\theta}\;D_{i}\phi \qquad \qquad B_{i}=-\frac{1}{2}\epsilon^{ijk}F_{jk}=\cos{\theta}D_{i}\phi
$$
with $\theta$ an arbitrary constant angle. 

At spatial infinity ($r\rightarrow \infty$, $\hat{r}^{2}=1$), we have
$$
D^{i}\phi \rightarrow \frac{\hat{r}^{i}}{4\pi r^{2}}G(\hat{r}),
$$ 
where $G(\hat{r})$ is an element of the Lie algebra of $H$, covariantly constant $D_{\mu}G(\hat{r})=0$\cite{olive-manton}. Also, the field strength goes to zero there, which leads to a asymptotically flat gauge field
$$
A_{\mu}=\frac{i}{e}\pa_{\mu}WW^{-1}
$$
and therefore, at $\mathcal{S}^{2,(t)}_{\infty}$ we get
$$
G\({\hat{r}}\)=WG_{R}W^{-1}
$$
where $G_{R}$ stands for the value of $G(\hat{r})$ at $x_{R}$.

Then, the operator $V(\Omega)$ discussed previously becomes
$$
V_{x_R^{(t)}}(\Omega_{\infty}^{(t)})=P_2 e^{ie\int_{S^{2,(t)}_{\infty}}d\tau d\sigma \left[ \alpha F_{\mu\nu}^W+\beta {\widetilde F}_{\mu\nu}^W\right] \frac{dx^{\mu}}{d\sigma}\frac{dx^{\nu}}{d\tau}}=e^{\left[-ie(\alpha\,\cos \theta+\beta\,\sin\theta)G_R\right]}
$$
and the conserved charges are given by the eigenvalues of $G_{R}$, which contain the magnetic and electric charges of the dyon solution. It is important to remark that since $G_{R}$ is at spatial infinity, according to our observations a change of the reference point makes it change by conjugation with the Wilson line of the path between the new point and the old one; this changes nothing in the eigenvalues of $G_{R}$.

\vspace{1cm}

{\bf Acknowledgements} The authors are grateful to fruitful discussions with O. Alvarez, E. Castellano, P. Klimas, M.A.C. Kneipp, R. Koberle, J. S\'anchez-Guill\'en, N. Sawado and W. Zakrzewski. LAF is partially supported by CNPq, and GL is supported by a CNPq scholarship.

\newpage


\appendix

\noindent{\Large {\bf Appendices}}

\vspace{.5cm}

We give in the next two appendices the proofs,  following the arguments of \cite{afs,afs-review}, of the  standard non-abelian Stokes theorem for a $1$-form connection and then its generalization for a $2$-form connection. 
 
\section{The ``standard'' non-abelian Stokes theorem}
\label{sec:curves}
\setcounter{equation}{0}

Consider a Lie algebra valued 1-form $C=C_{\mu}(x)dx^{\mu}$, defined in a $d+1$ dimensional simply connected space-time $M$. Given a path $\gamma$ parametrized by $\sigma \in [0,2\pi]$, such that $x^{\mu}(\sigma=0)\equiv x_{R}$ and $x^{\mu}(\sigma = 2\pi )\equiv x_{f}$, the Wilson line is constructed from\footnote{The notation $W_{\gamma}[\sigma ,0]$ says that the quantity $W$ is evaluated on the path $\gamma$ from the point with $\sigma=0$ to the point $x^{\mu}(\sigma)$.} 
\begin{equation}
\label{holo_w}
\frac{dW_{\gamma}[\sigma ,0]}{d\sigma}+C_{\mu}(x)\frac{dx^{\mu}}{d\sigma}W_{\gamma}[\sigma ,0]=0,
\end{equation}
with initial condition $W[0,0]=W_{R}$. 

After integrated, equation \eqref{holo_w} gives the path-ordered integral
\begin{equation}
\label{w_sigma}
W_{\gamma}[\sigma ,0] =P_{1}\;\exp{\lp - \int_{0}^{\sigma}C_{\mu}\frac{dx^{\mu}}{d\sigma^{\prime}}d\sigma^{\prime}   \rp}\cdot W_{R}.
\end{equation}
There are many paths one can choose to link $x_{R}$ to $x_{f}$, and a natural question is whether the Wilson line depends upon this choice. Being $M$ simply connected, a different path $\gamma^{\prime}$ with the same end points $x_{R}$ and $x_{f}$ is related to $\gamma$ by continuous transformations and the Wilson line reads $W_{\gamma^{\prime}}[2\pi,0]$.

Lets introduce a new parameter, $\tau \in [0,2\pi]$, labeling the path one chooses to go from $x_{R}$ to $x_{f}$. For $\gamma$ we set $\tau=0$, and for $\gamma^{\prime}$, $\tau=2\pi$. All other values give intermediary paths between these two, that arise due to the variation process $\gamma \rightarrow \gamma + \delta \gamma$.
With that, each point is now characterized by two parameters, $\tau$ and $\sigma$: the first tells in which curve the point is, while the second, where in this curve.

The variation is performed as follows. At a given point, immediately after $x_{R}$ we define a vector $T^{\mu}=\frac{dx^{\mu}}{d\tau}$, joining $\gamma$ and $\gamma+\delta \gamma$. Basically we are defining a way to map one point of the curve at $\tau=0$ to a point of the curve at $\tau\neq 0$. In order to answer if this holds for every other point, we need to parallel transport the vector $T^{\mu}$ along $\gamma$, \emph{i.e.}, along the direction of $S^{\mu}=\frac{dx^{\mu}}{d\sigma}$; the Lie derivative of $T^{\mu}$ in the direction of $S^{\mu}$ gives
$
\mathfrak{L}_{S}T^{\mu}=S^{\nu}\pa_{\nu}T^{\mu}-T^{\nu}\pa_{\nu}S^{\mu}=\frac{d^{2}x^{\mu}}{d\sigma d\tau}-\frac{d^{2}x^{\mu}}{d\tau d\sigma }=0
$
and therefore if we define the variation to be orthogonal to the curve $\gamma$ at a given point, this will be the case for every other point.
\begin{figure}  
  \begin{center}
    \includegraphics[width=0.3\textwidth]{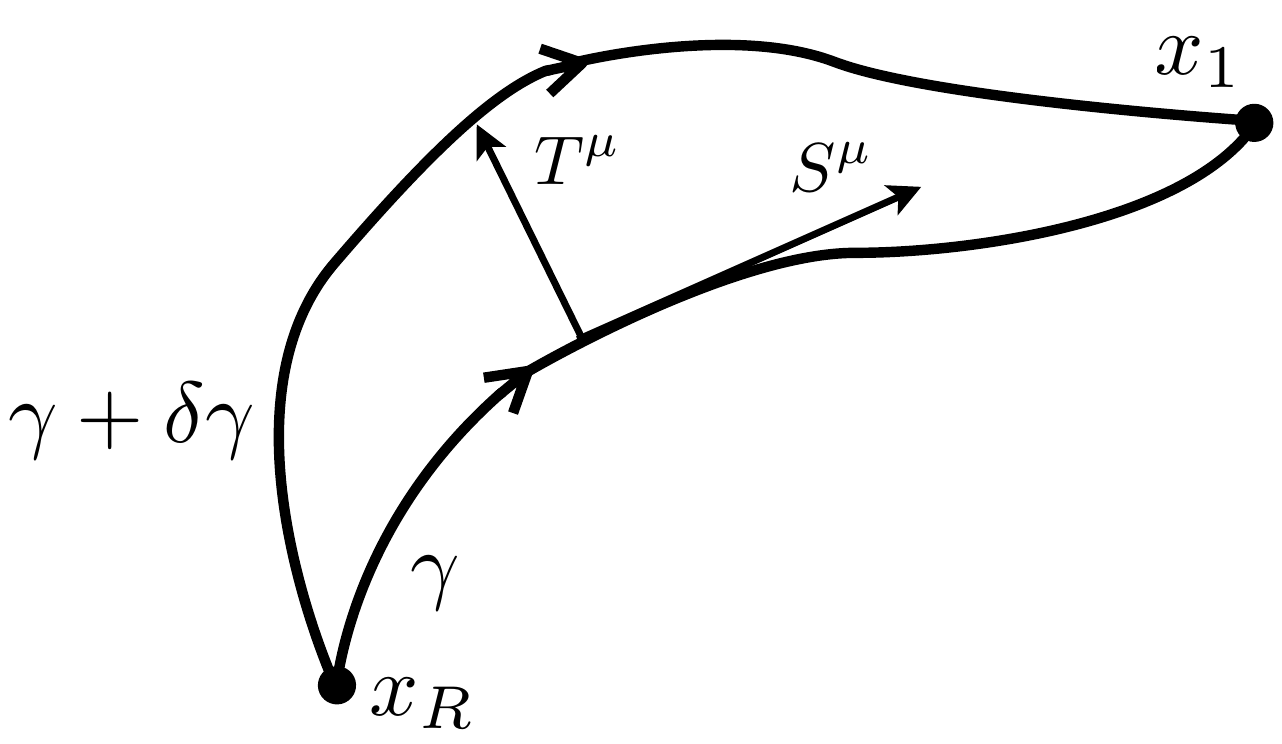}
\caption{The variation of a path with fixed end points.}  
  \end{center}
  \end{figure}
A variation of the path implies a variation of the holonomy, which can be calculated taking the variation of the equation \eqref{holo_w}:
$$
\delta \lp \frac{dW_{\gamma}}{d\sigma}+C_{\mu}(x)\frac{dx^{\mu}}{d\sigma}W_{\gamma}\rp =0  .
$$
Computing this explicitly\footnote{We introduced a new, but obvious, notation: $\delta W_{\gamma}[\tau;\sigma ,0]$. This stands for the variation of the Wilson line at a point $x^{\mu}(\sigma)$ in the curve $\tau$. } (omitting some symbols that we shall reintroduce at the end):
\begin{eqnarray*}
W^{-1}\frac{d}{d\sigma}\lp \delta W \rp + W^{-1}C_{\mu}\frac{dx^{\mu}}{d\sigma}\delta W+W^{-1}\delta \lp C_{\mu}\frac{dx^{\mu}}{d\sigma} \rp W \= 0\\
\frac{d}{d\sigma}\lp W^{-1}\delta W \rp -\lp \frac{d}{d\sigma}W^{-1} \rp \delta W + W^{-1}C_{\mu}\frac{dx^{\mu}}{d\sigma}\delta W+W^{-1}\delta \lp C_{\mu}\frac{dx^{\mu}}{d\sigma} \rp W \= 0\\
\frac{d}{d\sigma}\lp W^{-1}\delta W \rp + W^{-1}\delta \lp C_{\mu}\frac{dx^{\mu}}{d\sigma} \rp W \= 0
\end{eqnarray*} 
where after the variation was performed we multiplied the equation by $W^{-1}$ from the left (line 1), used the chain rule in order to rewrite the first term (line 2), and with the identity $\frac{dW^{-1}}{d\sigma}=-W^{-1}\frac{dW}{d\sigma}W^{-1}$ and equation \eqref{holo_w}, two terms were mutually canceled (line 3). 
This last equation can be integrated from $x^{\mu}(\sigma =0)$ to $x^{\mu}(\sigma)$, a point of $\gamma$. Taking into account the fact that the holonomy does not change at the initial point ($\delta W_{0}=0$), we get
$$
\delta W_{\gamma}[\sigma ,0]=-W_{\gamma}[\sigma,0]\;\int_{0}^{\sigma}d\sigma^{\prime}  W^{-1}_{\gamma}[\sigma^{\prime},0] \delta \lp C_{\mu}\frac{dx^{\mu}}{d\sigma^{\prime}} \rp W_{\gamma}[\sigma^{\prime},0].
$$
The final step is to compute the variation appearing inside the integral above:
\begin{eqnarray*}
&&\int_{0}^{\sigma}d\sigma^{\prime}\; W^{-1} \delta \lp C_{\mu}\frac{dx^{\mu}}{d\sigma^{\prime}} \rp W =\\
&& W^{-1}(\sigma)C_{\mu}\lp x\lp \sigma \rp \rp W(\sigma)\delta x^{\mu}(\sigma)+
\int_{0}^{\sigma}d\sigma^{\prime}\; \left\{ W^{-1}  \delta C_{\mu}\frac{dx^{\mu}}{d\sigma^{\prime}}  W  - \frac{d}{d\sigma^{\prime}}\lp W^{-1}C_{\mu}W \rp \delta x^{\mu}\right\}  = \\
&&W^{-1}C_{\mu} W\delta x^{\mu}+\int_{0}^{\sigma}d\sigma^{\prime}\; W^{-1}\left\{   \pa_{\nu}C_{\mu}\delta x^{\nu}\frac{dx^{\mu}}{d\sigma^{\prime}}    -  \frac{dC_{\mu}}{d\sigma^{\prime}}\delta x^{\mu}+\lc C_{\mu},\frac{dW}{d\sigma^{\prime}}W^{-1} \rc \delta x^{\mu} \right\}  W   =\\
&&W^{-1}C_{\mu} W\delta x^{\mu}+\int_{0}^{\sigma}d\sigma^{\prime}\; W^{-1}\left\{   \pa_{\nu}C_{\mu}\delta x^{\nu}\frac{dx^{\mu}}{d\sigma^{\prime}}    -  \pa_{\nu}C_{\mu}\frac{dx^{\nu}}{d\sigma^{\prime}}\delta x^{\mu}-\lc C_{\mu},C_{\nu}\frac{dx^{\nu}}{d\sigma^{\prime}} \rc \delta x^{\mu} \right\}  W   =\\
&&W^{-1}(\sigma)C_{\mu}\lp x\lp \sigma \rp \rp W(\sigma)\delta x^{\mu}(\sigma)-\int_{0}^{\sigma}d\sigma^{\prime}\; W^{-1}F_{\mu\nu}W\frac{dx^{\mu}}{d\sigma^{\prime}}\delta x^{\nu}
\end{eqnarray*}
where $F_{\mu\nu}=\pa_{\mu}C_{\nu}-\pa_{\nu}C_{\mu}+[C_{\mu} , C_{\nu}]$ is the curvature of $C_{\mu}$.

From line 1 to line 2 we performed an integration by parts and used the fact that $\delta x^{\mu}(0)=0$. Then, since the variation in the connection is due to a variation in the space-time point ($x \rightarrow x + \delta x$), we took $\delta C_{\mu}=\pa_{\nu}C_{\mu}\delta x^{\nu}$ in line 3, and performed the derivative. After that, we used chain rule for the derivative of the connection in $\sigma^{\prime}$, and equation \eqref{holo_w} inside the commutator, which gives line 4. Then, relabeling the indices we get a curvature, showed in line 5.

Finally, the variation of the Wilson line due to a variation of the path $\gamma$ to another path, labeled by $\tau$ reads
\begin{eqnarray}
\label{holo_w_var}
\delta W_{\gamma}[\tau;\sigma ,0]&=&-C_{\mu} W_{\gamma}[\tau;\sigma ,0]\delta x^{\mu}(\sigma)+\nonumber \\ 
&& W_{\gamma}[\tau;\sigma ,0]\int_{0}^{\sigma}d\sigma^{\prime}\; W^{-1}_{\gamma}[\tau;\sigma^{\prime},0]F_{\mu\nu}W_{\gamma}[\tau;\sigma^{\prime},0]\frac{dx^{\mu}}{d\sigma^{\prime}}\delta x^{\nu}
\end{eqnarray}

Notice that this is a general result for the variation is not specified: one can do it in both directions, tangent to the path, or orthogonal to it. Of course, in the tangent direction, a variation is just a reparametrization $\sigma \rightarrow \widetilde{\sigma}=f(\sigma)$ , which, from equation \eqref{holo_w}, changes nothing as long as $f(\sigma)$ is monotonic. On the other hand, in the orthogonal direction one has $\delta x^{\mu} = T^{\mu}$ and $\delta W[\tau]=W[\tau+\delta \tau]-W[\tau]=\frac{dW}{d\tau}\delta \tau$, which, in \eqref{holo_w_var} (with $\sigma = 2\pi$, so that the end points are the same for both paths, and therefore $\delta x^{\mu}(0)=\delta x^{\mu}(2\pi)=0$) gives a differential equation for $W$:
\begin{equation}
\label{holo_w_tau}
\frac{d}{d\tau} W_{\gamma}[\tau;2\pi ,0]- W_{\gamma}[\tau;2\pi ,0]\int_{0}^{2\pi}d\sigma\; W^{-1}_{\gamma}[\tau;\sigma,0]F_{\mu\nu}W_{\gamma}[\tau;\sigma,0]\frac{dx^{\mu}}{d\sigma}\frac{dx^{\nu}}{d\tau}=0.
\end{equation}
What we conclude is that there are two possible ways to calculate the Wilson line of a given curve. Consider, for instance, the path $\gamma^{\prime}$. The Wilson line $W_{\gamma^{\prime}}[\tau=2\pi;x_{f},x_{R}]$ can be obtained, first, from equation \eqref{holo_w}, after integration over $\gamma^{\prime}$. The second possibility is to think of $\gamma^{\prime}$ as  obtained from $\gamma$, using the variations we discussed above. Then the holonomy there is given after integrating \eqref{holo_w_tau} from $\tau=0$ to $\tau=2\pi$:
$$
W_{\gamma^{\prime}}[2\pi;2\pi,0]=W_{\gamma}[0;2\pi,0]\cdot P_{2}\;\exp{\lp \int_{0}^{2\pi}d\tau \int_{0}^{2\pi}d\sigma\; W^{-1}_{\gamma^{\prime}}F_{\mu\nu}W_{\gamma^{\prime}}\frac{dx^{\mu}}{d\sigma}\frac{dx^{\nu}}{d\tau}\rp}
$$
where $P_{2}$ stands for the ``surface-ordering'', \emph{i.e.}, the ordering due to the non-abelian character of the fields, according to the way we vary $\tau$.

It is clear that the quantity appearing in the r.h.s of the above equation is the flux of $F_{\mu\nu}^{W}\equiv W^{-1}F_{\mu\nu}W$ through the space-time surface $\Sigma \subset M$, whose boundary is $\partial \Sigma = \Gamma \equiv \gamma^{\prime -1}\cdot \gamma$; lets call it $\Phi(F^{W},\Sigma)$, and rewrite the equation as
\begin{equation}
W_{\gamma^{\prime}}=W_{\gamma}\cdot P_{2}\;e^{\Phi(F^{W},\Sigma)}.
\end{equation}
Consider the case where the path discussed above is a loop $\Gamma$ with no self intersections:
$$
x^{\mu}: \sigma \in [0,2\pi]\rightarrow \mathcal{M} \qquad \sigma \mapsto x^{\mu}(\sigma);\qquad x^{\mu}(0)=x^{\mu}(2\pi).
$$
Then, integration of \eqref{holo_w} for the whole loop gives
\begin{equation}
W_{\Gamma}=P_{1}\;e^{\oint_{\Gamma}C_{\mu}dx^{\mu}}\cdot W_{R}.
\end{equation}
Being $\Gamma$ the boundary of a two dimensional submanifold $\Sigma \subset M$, we also have that this same Wilson line can be calculated taking a variation from the point loop $P_{x_{R}}=x_{R}\; \forall \; \sigma \in [0,2\pi]$ (whose Wilson line is $W_{R}$)
\begin{equation}
W_{\Gamma}=W_{R}\cdot P_{2}\;e^{\Phi(F^{W},\Sigma)}.
\end{equation}
The fact that it is possible to calculate $W_{\Gamma}$ integrating the connection $C$ over $\Gamma$, or integrating the curvature $F(C)$ over the area bounded by $\Gamma$ is exactly the statement of the Stokes theorem:
\begin{equation}
\label{stokes_w}
P_{1}\;\exp{\lp \oint_{\partial \Sigma} C_{\mu}dx^{\mu}\rp}\cdot W_{R}=W_{R}\cdot P_{2}\;\exp{\lp \Phi(F^{W},\Sigma)\rp}.
\end{equation}

\section{A generalization of the non-abelian Stokes theorem for a 2-form connection}
\label{sec:surface}
\setcounter{equation}{0}

Let us consider the antisymmetric field\footnote{The 2-form $B=\frac{1}{2}B_{\mu\nu}(x)dx^{\mu}\wedge dx^{\nu}$ is not necessarily exact.} $B_{\mu\nu}(x)$, defined in the $d+1$ dimensional space-time $M$. A family of (homotopically equivalent) loops with base point $x_{R}$ can be used to scan the two-dimensional hypersurface $\Sigma$, starting from the infinitesimal loop around $x_{R}$ and taking variations along the $T^{\mu}$ direction, as explained in the previous section.

A new quantity\footnote{For future reference we call it surface-holonomy.} $V$ is introduced in analogy with the Wilson line in equation \eqref{holo_w_tau}, defined by
\begin{equation}
\label{holo_v}
\frac{dV_{\Sigma}[\tau ,0]}{d\tau} - V_{\Sigma}[\tau ,0]\int_{0}^{2\pi}d\sigma\; W^{-1}_{\Gamma}[\tau;\sigma,0]B_{\mu\nu}W_{\Gamma}[\tau;\sigma,0]\frac{dx^{\mu}}{d\sigma}\frac{dx^{\nu}}{d\tau}=0,
\end{equation}
with the initial condition $V[0,0]\equiv V_{\Sigma_{R}}$, being $\Sigma_{R}$ the infinitesimal surface around $x_{R}$.

The surface-holonomy $V_{\Sigma}[\tau,0]$ is defined on the surface whose boundary is the loop labeled by $\tau$. The quantity
$$
T_{2\pi}(B,C,\tau)\equiv \int_{0}^{2\pi}d\sigma\; W^{-1}_{\Gamma}[\tau;\sigma,0]B_{\mu\nu}W_{\Gamma}[\tau;\sigma,0]\frac{dx^{\mu}}{d\sigma}\frac{dx^{\nu}}{d\tau}
$$
plays the role of a ``non-local connection''\footnote{It was shown in \cite{afs, afs-review} that it is in fact a 1-form connection in the loop space.}, defined on each loop. The Wilson lines appearing inside the integral are calculated from \eqref{holo_w}, running from $x_{R}$ to $x^{\mu}(\sigma)$ on the loop $\tau$.

After integrating \eqref{holo_v} one gets the surface-ordered integral
\begin{equation}
V_{\Sigma}[\tau,0]=V_{R}\cdot P_{2}\;\exp{\lp\int_{0}^{\tau}d\tau^{\prime}\;T_{2\pi}(B,C,\tau^{\prime})\rp}.
\end{equation}
In analogy with the fact that the Wilson line is defined over the path that links two (boundary) points, the surface-holonomy is defined over the surface that links two (boundary) loops.  Take these two loops to be those at $\tau=0$ and $\tau=2\pi$, fixed. Clearly one can use different surfaces to link them. Consider $\Sigma$ and $\Sigma^{\prime}$, two possibilities.
Each of these choices might lead to a different solution of \eqref{holo_v}. Since $\Sigma^{\prime}$ can be obtained from $\Sigma$ after a variation in the $Z^{\mu}\equiv \frac{dx^{\mu}}{d\zeta}$ direction\footnote{This direction is in the normal direction to the surface $\Sigma$.} ( $\zeta \in [0,2\pi]$ ) we can calculate the difference $\delta V = V[\zeta+\delta \zeta]-V[\zeta]=\frac{dV}{d\zeta}\delta \zeta$ in analogy with what we did for the path-holonomy, taking first the variation of the defining equation \eqref{holo_v}:
$$
\delta \lp \frac{dV_{\Sigma}[\tau ,0]}{d\tau} - V_{\Sigma}[\tau ,0]T_{2\pi}(B,C,\tau) \rp=0
$$
Making it explicitly:
\begin{eqnarray*}
\frac{d}{d\tau}\lp \delta VV^{-1} \rp - \delta V\lp \frac{dV^{-1}}{d\tau}+ T_{2\pi}(B,C,\tau)V^{-1} \rp - V\delta T_{2\pi}(B,C,\tau)V^{-1}\= 0 \\
\frac{d}{d\tau}\lp \delta VV^{-1} \rp  - V\delta T_{2\pi}(B,C,\tau)V^{-1}\= 0 
\end{eqnarray*}
We took the variation, and multiplied by $V^{-1}$ from the right. After that the chain rule was used to get the first two terms in line 1. Then, using the identity $\frac{dV^{-1}}{d\tau}=-V^{-1}\frac{dV}{d\tau}V^{-1}$ and equation \eqref{holo_v} the term inside the parenthesis in line 2 vanishes. The next step is to expand the term $VT_{2\pi}(B,C,\tau)V^{-1}$ appearing above.
\begin{eqnarray*}
&&V\delta T_{2\pi}(B,C,\tau)V^{-1}= V \int_{0}^{2\pi} d\sigma \lp \delta W^{-1}B_{\mu\nu}W + WB_{\mu\nu}\delta W^{-1} + W \delta B_{\mu\nu} W^{-1} \rp \frac{dx^{\mu}}{d\sigma}\frac{dx^{\nu}}{d\tau}V^{-1}\\
&+&V\int_{0}^{2\pi} d\sigma \; W^{-1}B_{\mu\nu}W \frac{d\delta x^{\mu}}{d\sigma}\frac{dx^{\nu}}{d\tau}V^{-1} + V\int_{0}^{2\pi} d\sigma \; W^{-1}B_{\mu\nu}W \frac{dx^{\mu}}{d\sigma}\frac{d\delta x^{\nu}}{d\tau} V^{-1} \\
&=& V \int_{0}^{2\pi} d\sigma \lp \lc B^{W}_{\mu\nu},W^{-1}\delta W \rc + W^{-1}\pa_{\rho}B_{\mu\nu}W\delta x^{\rho}  \rp\frac{dx^{\mu}}{d\sigma}\frac{dx^{\nu}}{d\tau}V^{-1}\\
&-&V\int_{0}^{2\pi}d\sigma\;\lp \lc B^{W}_{\mu\nu},W^{-1}\frac{dW}{d\sigma} \rc\frac{dx^{\nu}}{d\tau}+W^{-1}\pa_{\rho}B_{\mu\nu}W\frac{dx^{\rho}}{d\sigma}\frac{dx^{\nu}}{d\tau}+ B^{W}_{\mu\nu}\frac{d^{2}x^{\nu}}{d\sigma d\tau} \rp\delta x^{\mu}V^{-1}\\
&+& \frac{d}{d\tau}\lp V T_{2\pi}(B,C,\delta)V^{-1}\rp - \frac{dV}{d\tau}T_{2\pi}(B,C,\delta)V^{-1}-VT_{2\pi}(B,C,\delta)\frac{dV^{-1}}{d\tau}\\
&-&V\int_{0}^{2\pi} d\sigma \; \frac{d}{d\tau}\lp W^{-1}B_{\mu\nu}W \frac{dx^{\mu}}{d\sigma}\rp V^{-1}\delta x^{\nu}.
\end{eqnarray*}
When the integration by parts of the second term was performed we used the fact that $\delta x^{\mu}(\sigma)$ vanishes at the boundaries. 
For the commutators we use $\delta W$ given by \eqref{holo_w_var} and $\frac{dW}{d\sigma}$ given by \eqref{holo_w}. In line 5 we can use the identity $\frac{dV^{-1}}{d\tau}=-V^{-1}\frac{dV}{d\tau}V^{-1}$ and we calculate the derivative in the last term using equation \eqref{holo_v}.

Plugging the result back into the equation we started with, and integrating gives
\begin{eqnarray*}
\delta V_{\Sigma}[\tau,0]\=   \lp V T_{2\pi}(B,C,\delta)V^{-1} \rp \Bigg\vert^{\tau}_{0}V_{\Sigma}[\tau ,0]+\\
&+&\int_{0}^{\tau} d\tau^{\prime} V\Bigg\{
\int_{0}^{2\pi}d\sigma W^{-1}\lp D_{\rho}B_{\mu\nu}+D_{\mu}B_{\nu\rho}+D_{\nu}B_{\rho\mu} \rp W\frac{dx^{\mu}}{d\sigma}\frac{dx^{\nu}}{d\tau^{\prime}}\delta x^{\rho}+\\
&+&\int_{0}^{2\pi}d\sigma \lc B_{\mu\nu}^{W},T_{\sigma}(F,C,\delta) \rc \frac{dx^{\mu}}{d\sigma}\frac{dx^{\nu}}{d\tau^{\prime}}
-\int_{0}^{2\pi}d\sigma \lc B_{\mu\nu}^{W},T_{\sigma}(F,C,\tau^{\prime}) \rc \frac{dx^{\mu}}{d\sigma}\delta x^{\nu}+\\
&+&\lc T_{2\pi}(B,C,\delta),T_{2\pi}(B,C,\tau^{\prime})\rc 
\Bigg\}V^{-1}\;
V_{\Sigma}[\tau ,0].
\end{eqnarray*}
The second term will be called $\mathcal{K}$, so that we can write
\begin{equation}
\label{holo_v_var}
\delta V_{\Sigma}[\tau,0] = \lp V T_{2\pi}(B,C,\delta)V^{-1} \rp \Bigg\vert^{\tau}_{0}V_{\Sigma}[\tau ,0] + \mathcal{K} V_{\Sigma}[\tau,0].
\end{equation}
Taking $\tau=2\pi$, the first term on the RHS vanishes and we get
\begin{equation}
\label{close_holo_v_var}
\delta V_{\Sigma}[2\pi,0] - \mathcal{K} V_{\Sigma}[2\pi,0]=0.
\end{equation}
Then, taking the variation to be along the $Z^{\mu}$ direction so that $\delta V = \frac{dV}{d\zeta}\delta \zeta$, we get the differential equation for $V$(now we drop some of the symbols that have being used so far):
\begin{equation}
\label{holo_v_zeta_2}
\frac{d}{d\zeta}V-\mathcal{K} V=0,
\end{equation}
where all $\delta x^{\mu}$ in $\mathcal{K}$ is replaced by $\frac{dx^{\mu}}{d\zeta}\delta \zeta$, and we write it in the nicer form:
\begin{eqnarray*}
\mathcal{K}\=\int_{0}^{2\pi}d\tau V\Bigg\{\int_{0}^{2\pi}d\sigma\;W^{-1}\lp D_{\rho}B_{\mu\nu}+D_{\mu}B_{\nu\rho}+D_{\nu}B_{\rho\mu} \rp W\frac{dx^{\mu}}{d\sigma}\frac{dx^{\nu}}{d\tau}\frac{dx^{\rho}}{d\zeta}+\\
&-&\int_{0}^{2\pi}d\sigma \int_{0}^{\sigma}d\sigma^{\prime}\;\lc B^{W}_{\kappa \lambda}(\sigma^{\prime})-F^{\mu\nu}_{\kappa \lambda}(\sigma^{\prime}),B^{W}_{\mu\nu}(\sigma) \rc \frac{dx^{\kappa}}{d\sigma^{\prime}}(\sigma^{\prime})\frac{dx^{\mu}}{d\sigma}(\sigma)\times \\
&\times &\lp  \frac{dx^{\lambda}}{d\tau}(\sigma^{\prime})\frac{dx^{\nu}}{d\zeta}(\sigma)-
\frac{dx^{\lambda}}{d\zeta}(\sigma^{\prime})\frac{dx^{\nu}}{d\tau}(\sigma)  
\rp\Bigg\}V^{-1}.
\end{eqnarray*}
Now let us consider a closed surface $\Sigma_{\circ}$, so that the solution of the equation \eqref{holo_v} reads
\begin{equation}
\label{v_tau}
 V_{\Sigma_{\circ}}=V_{R}\cdot P_{2}\;\exp{\lp\int_{0}^{2\pi}d\tau\;T_{2\pi}(B,A,\tau)\rp}.
\end{equation}
The surface $\Sigma_{\circ}$ is the boundary of the three-dimensional submanifold (volume) $\Omega$, so the surface-holonomy $V_{\Sigma_{\circ}}$ can be calculated from variations in the $Z^{\mu}$ direction starting at the infinitesimal surface $\Sigma_{R}$ around the reference point $x_{R}$, using \eqref{holo_v_zeta_2}:
\begin{equation}
\label{v_zeta}
V_{\Sigma_{\circ}}=P_{3}\;\exp{\lp \int_{0}^{2\pi}d\zeta\; \mathcal{K}\rp}V_{\Sigma_{R}}.
\end{equation}
The fact that there are two ways to compute the surface-holonomy $V_{\Sigma_{\circ}}$ is the statement of the Stokes theorem:
\begin{equation}
\label{stokes_v}
V_{\Sigma_{R}}P_{2}\;\exp{\lp\int_{0}^{2\pi}d\tau\;T_{2\pi}(B,A,\tau)\rp}=P_{3}\;\exp{\lp \int_{0}^{2\pi}d\zeta\; \mathcal{K}\rp}V_{\Sigma_{R}}
\end{equation}
being the l.h.s computed over the surface $\Sigma$ which is the boundary of the volume $\Omega$, where we integrate the r.h.s.
In order to make it more explicit we rewrite this as
$$
V_{\Sigma_{R}}P_{2}\;\exp{\lp\int_{\partial \Omega }d\tau\;d\sigma\;W^{-1}B_{\mu\nu}W\frac{dx^{\mu}}{d\sigma}\frac{dx^{\nu}}{d\tau}\rp}=P_{3}\;\exp{\lp \int_{0}^{2\pi}d\zeta\; \mathcal{K}\rp}V_{\Sigma_{R}}.
$$

\end{document}